\documentclass[fleqn,usenatbib]{mnras}
\usepackage[T1]{fontenc}
\DeclareRobustCommand{\VAN}[3]{#2}
\let\VANthebibliography\thebibliography
\def\thebibliography{\DeclareRobustCommand{\VAN}[3]{##3}\VANthebibliography}
\usepackage{graphicx,siunitx,xcolor,tikz, upquote,bm}	
\usetikzlibrary{decorations.markings,arrows}
\usepackage{amsmath,lmodern,anyfontsize}	
\usepackage{amssymb}	
\usepackage{fancyvrb}
\usepackage{subcaption,soul}
\graphicspath{{figs/}}

\makeatletter
\newcommand{\cz}{
  \mathord{\mathpalette\vaggelis@z{z}}%
}
\newcommand{\cZ}{
  \mathord{\mathpalette\vaggelis@z{Z}}%
}
\newcommand{\vaggelis@z}[2]{%
  \sbox\z@{$\m@th#1#2$}%
  \ooalign{%
    $\m@th#1#2$\cr
    \hidewidth
    \vrule height \dimexpr.5\ht\z@+0.03ex\relax
           depth -\dimexpr.5\ht\z@-0.03ex\relax
           width .5\wd\z@
    \hidewidth\cr
  }%
  \vphantom{\box\z@}
}
\makeatother

\newcommand{\ud}{\mathrm{d}}   
\newcommand{\ue}{\mathrm{e}} 
\definecolor{lime}{HTML}{A6CE39}
\DeclareRobustCommand{\orcidicon}{\hspace{-3mm}
	\begin{tikzpicture}
	\draw[lime, fill=lime] (0,0) 
	circle [radius=0.16] 
	node[white] {\hspace{0.1mm}{\fontfamily{qag}\selectfont \tiny ID}};
	\draw[white, fill=white] (-0.07,0.1) 
	circle [radius=0.01];
	\end{tikzpicture}
	\hspace{-5mm}
}

\foreach \x in {A, ..., Z}{\expandafter\xdef\csname orcid\x\endcsname{\noexpand\href{https://orcid.org/\csname orcidauthor\x\endcsname}
		{\noexpand\orcidicon}}
}
\DeclareSIUnit\parsec{pc}
\defcitealias{Zheng}{CZ07}
\defcitealias{Semelin}{S07}
\defcitealias{Loeb_1999}{LR99}

\title[Ly$\alpha$ RT at cosmic dawn]{Radiative transfer of Lyman-$\bm{\alpha}$ photons at cosmic dawn with realistic gas physics}

\author[Mittal et al.]{
Shikhar Mittal$^{1,2}$\,\ \orcidA{}\ \ \thanks{E-mail: sm2941@cam.ac.uk}, Girish Kulkarni$^{1}$\,\ \orcidB{}\ \ and Thibault Garel$^{3}$
\\
$^1$Tata Institute of Fundamental Research, Homi Bhabha Road, Mumbai 400005, India\\
$^{2}$Battcock Centre for Experimental Astrophysics, Cavendish Laboratory, J.~J.\ Thomson Avenue, Cambridge CB3 0HE, UK\\
$^3$Observatoire de Gen\`{e}ve, Universit\'{e} de Gen\`{e}ve, 51 Chemin Pegasi, CH-1290 Versoix, Switzerland
}

\date{Accepted 2024 October 30. Received 2024 October 10; in original form 2023 November 6}

\pubyear{2024}

\begin{document}
\setstcolor{red}
\VerbatimFootnotes
\label{firstpage}
\pagerange{\pageref{firstpage}--\pageref{lastpage}}
\maketitle

\begin{abstract}
Lyman-$\alpha$ photons enable the cosmic dawn 21-cm signal through a process called the Wouthuysen-Field effect. An accurate model of the signal in this epoch hinges on the accuracy of the computation of the Ly$\alpha$ coupling, which requires one to calculate the specific intensity of Ly$\alpha$ photons emitted from the first stars. Most traditional calculations of the Ly$\alpha$ coupling assume a delta-function scattering cross-section, as the resonant nature of the Ly$\alpha$ scattering makes an accurate radiative transfer (RT) solution computationally expensive. Attempts to improve upon this traditional approach using numerical RT have recently emerged. However, some of these treatments suffer from assumptions such as a uniform gas distribution, coherent scattering in the gas frame and isotropic scattering. While others which do not account for these only do so through certain schemes along with core-skipping algorithms. We present results from a self-consistent Monte Carlo RT simulations devoid of any of the assumptions in the previous work for the first time. We find that gas bulk motion is the most important effect to account for in RT resulting in an RMS difference of 38\% in the 21-cm signal and anisotropic scattering being the least important effect contributing to less than 3\% RMS difference in 21-cm signal. We also evaluate the 21-cm power spectrum and compare that with the traditional results at cosmic dawn. This work points the way towards higher-accuracy models to enable better inferences from future measurements.
\end{abstract}



\section{Introduction}

Before the emergence of the first stars, during the cosmic dark ages collisions of hydrogen atoms with each other and with other species bring hyperfine transitions of neutral hydrogen in equilibrium with the gas. This enables a global 21-cm signal in absorption, with the strongest feature being about $\SI{50}{\milli\kelvin}$ at $z\sim80$ \citep{Pritchard_2012}. As the Universe cools down and dilutes due to adiabatic Hubble expansion, the hyperfine transitions equilibriate with the cosmic microwave background (CMB). This washes out any contrast against the background, rendering the global 21-cm signal zero. Formation of the first stars marks the beginning of the cosmic dawn as they produce Lyman-series photons that tend to bring hyperfine transitions again in equilibrium with the gas. At the same time, radiation (mostly X-rays) heats up the gas \citep{Santos_2008, Baek10, Mesinger_13}. The interplay of these processes is expected to result in an absorption signal with an amplitude of the order of $\SI{100}{\milli\kelvin}$ at redshifts $z=15$--$20$.

Several global 21-cm experiments are in development or have been developed which target the cosmic dawn, such as the Experiment to Detect the Global EoR Signal \citep[\textit{EDGES},][]{Bowman}, Shaped Antenna measurement of the background RAdio Spectrum \citep[\textit{SARAS},][]{saras3}, Large Aperture Experiment to Detect the Dark Ages \citep[\textit{LEDA},][]{Bernardi_2015, Bernardi_16, Price}, Probing Radio Intensity at high-Z from Marion \citep[\textit{PRIzM},][]{philip}, and Radio Experiment for the Analysis of Cosmic Hydrogen \citep[\textit{REACH},][]{Eloy, reach}. At the same time, interferometers such as LOw Frequency ARray \citep[\textit{LOFAR},][]{LOFAR}, Murchison Widefield Array \citep[\textit{MWA},][]{mwa}, Square Kilometre Array \citep[\textit{SKA},][]{SKA}, The Amsterdam–ASTRON Radio Transients Facility and Analysis Center \citep[\textit{AARTFAAC},][]{aartfaac}, Hydrogen Epoch Reionization Array \citep[\textit{HERA},][]{HERA}, and New extension in Nan\c{c}ay upgrading LOFAR \citep[\textit{NenuFAR},][]{nenufar} are trying to measure 21-cm power spectrum. Despite the ongoing efforts on theoretical and experimental fronts we still lack a coherent picture of the cosmic dawn. Given the heightened experimental activity, improvements in the accuracy of the theoretical modelling of the cosmic dawn 21-cm signal are timely.

In this work, we focus on the accuracy of one particular aspect of 21-cm signal calculation: the Lyman-$\alpha$ (Ly$\alpha$) coupling caused by the Wouthuysen--Field (WF) effect \citep{Field, Wouth}. This refers to a change in the occupation number of hyperfine states due to resonance scattering of Ly$\alpha$ photons by the hydrogen atom. This effect makes the 21-cm signal distinguishable from the CMB. We study the radiative transfer (RT) of Ly$\alpha$ photons to understand the rate of scattering in a cosmological volume. Previous analytical global 21-cm signal calculations \citep[e.g.][]{FP06} or even semi-numerical ones \citep[e.g.][]{Mesinger_11} assumed that photons originating on the blue side of the Ly$\alpha$ line centre stream freely until they cosmologically redshift down to the line centre, where they are absorbed by the hydrogen atom and possibly produce a hyperfine transition. However, in reality Ly$\alpha$ photons have a large optical depth in the intergalactic medium (IGM) because of which they scatter multiple times even before they reach the Ly$\alpha$ frequency. Stated differently, the photons have a broad line profile because of which they can scatter on either side of the line centre.

Previous authors who improved this picture include \citet{Zheng}, \citet{Naoz}, and \citet{Reis_2021, Reis_2022}. The RT technique followed by these authors was based on the analytical treatment developed by \citet[][hereafter \citetalias{Loeb_1999}]{Loeb_1999}, where the intergalactic medium was considered uniform, homogeneous, neutral, and undergoing Hubble expansion at a zero temperature. Such an approach does not capture the inhomogeneity of the density and temperature of the intergalactic gas. \citet{Baek2009} were the first to study the 3D RT of Ly$\alpha$ photons on a real cosmological temperature and density distribution using the technology introduced by \citet[][hereafter \citetalias{Semelin}]{Semelin}. In their procedure they adopt core-skipping algorithms to speed up the RT simulations but use the pre-derived form  of spectral distortions given by \citet{Hirata} to account for temperature and density inhomogeneities. Still, in such a scheme the gas velocity effects are missed in the line core where the largest number of scatterings happen. Most recently \citet{Semelin2023} also use core-skipping algorithm but introduce a correction to the number of core scatterings based on local bulk velocity.

Solving the full RT equation coupled to the hydrodynamics is a computationally expensive task because of the high dimensionality of the RT equation and the huge difference in the timescales of RT and the hydrodynamics. But even without the hydrodynamics RT can be quite challenging. For this reason one of the most popular approaches for RT is a Monte Carlo (MC) technique, where one does not directly deal with any integro-differential equations but rather tracks individual photons. In this work, we use the Monte Carlo code \texttt{RASCAS} \citep{rascas} for 3D RT of Ly$\alpha$ photons for the epoch of cosmic dawn. 

This paper is organized as follows. In Section~\ref{theory}, we discuss the basics of 21-cm signal in brief. Our focus will be on the details of our Ly$\alpha$ RT for the computation of Ly$\alpha$ coupling. We present our results in Section~\ref{results}. In the same section we contrast our work with previous literature. We highlight caveats and discuss future work in Section~\ref{caveats}, and end with a summary in Section~\ref{Conc}. We use the following cosmological parameters: $\Omega_{\mathrm{m}}= 0.315$, $\Omega_{\mathrm{b}}=0.049$, $\Omega_\Lambda = 0.685$, $h=0.674$, $Y_{\mathrm{p}}=0.245$, $T_0=\SI{2.725}{\kelvin}$, $\sigma_8 = 0.811$ and $n_{\mathrm{s}} = 0.965$ \citep{Fixsen_2009, Planck}, where $T_0$ and $Y_{\mathrm{p}}$ are the CMB temperature measured today and primordial helium fraction by mass, respectively. We prefix the distance units with a `c' to indicate a comoving length while no prefix to indicate proper physical lengths. We use `$\cz$' and `$z$' to represent the Cartesian coordinate of the `$xyz$' system and cosmological redshift, respectively.

\section{Theory and Methods}\label{theory}
\begin{figure*}
\centering
\begin{subfigure}{0.48\textwidth}
\includegraphics[width=1\linewidth]{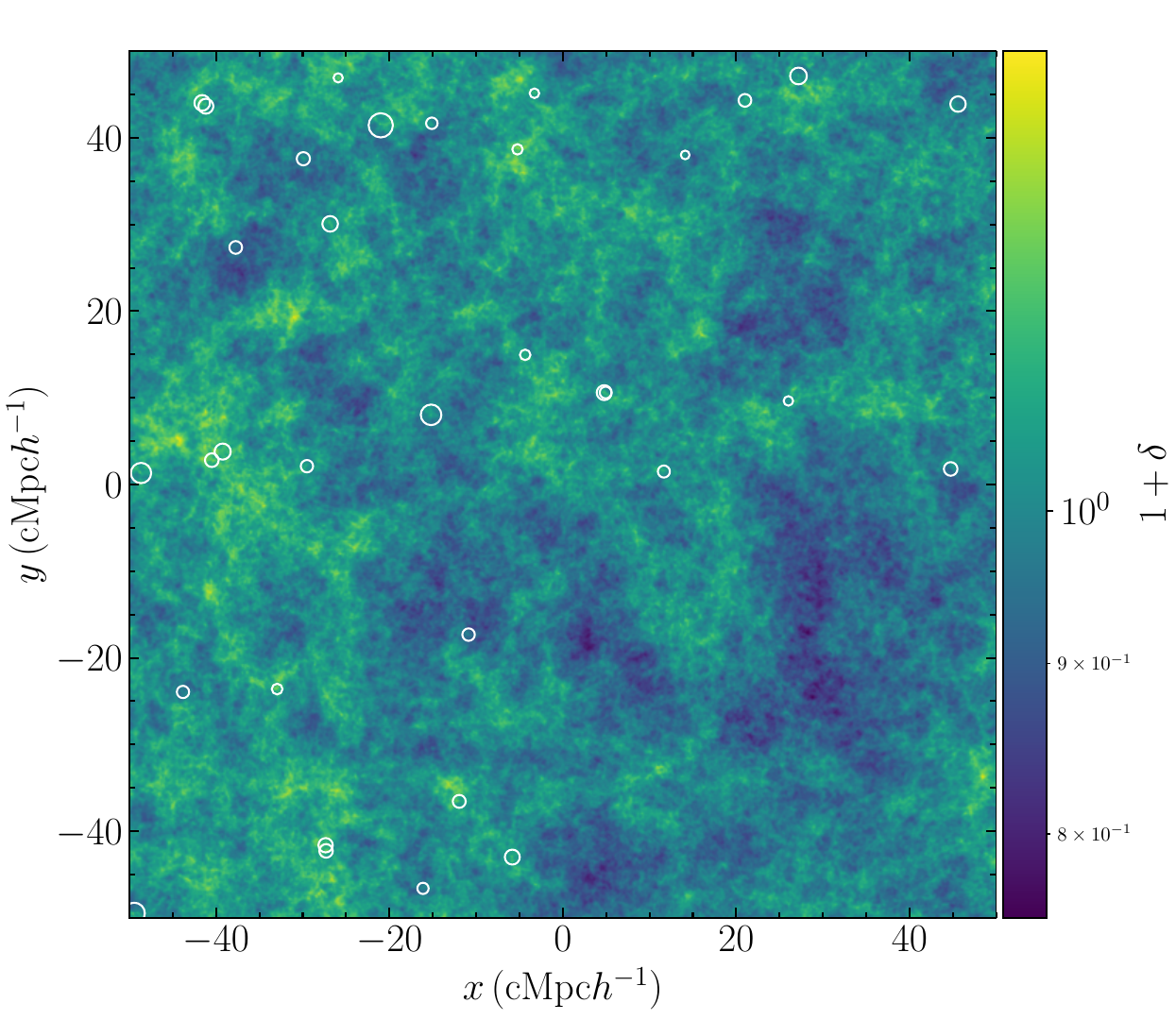}
\end{subfigure}%
\begin{subfigure}{0.48\textwidth}
\includegraphics[width=1\linewidth]{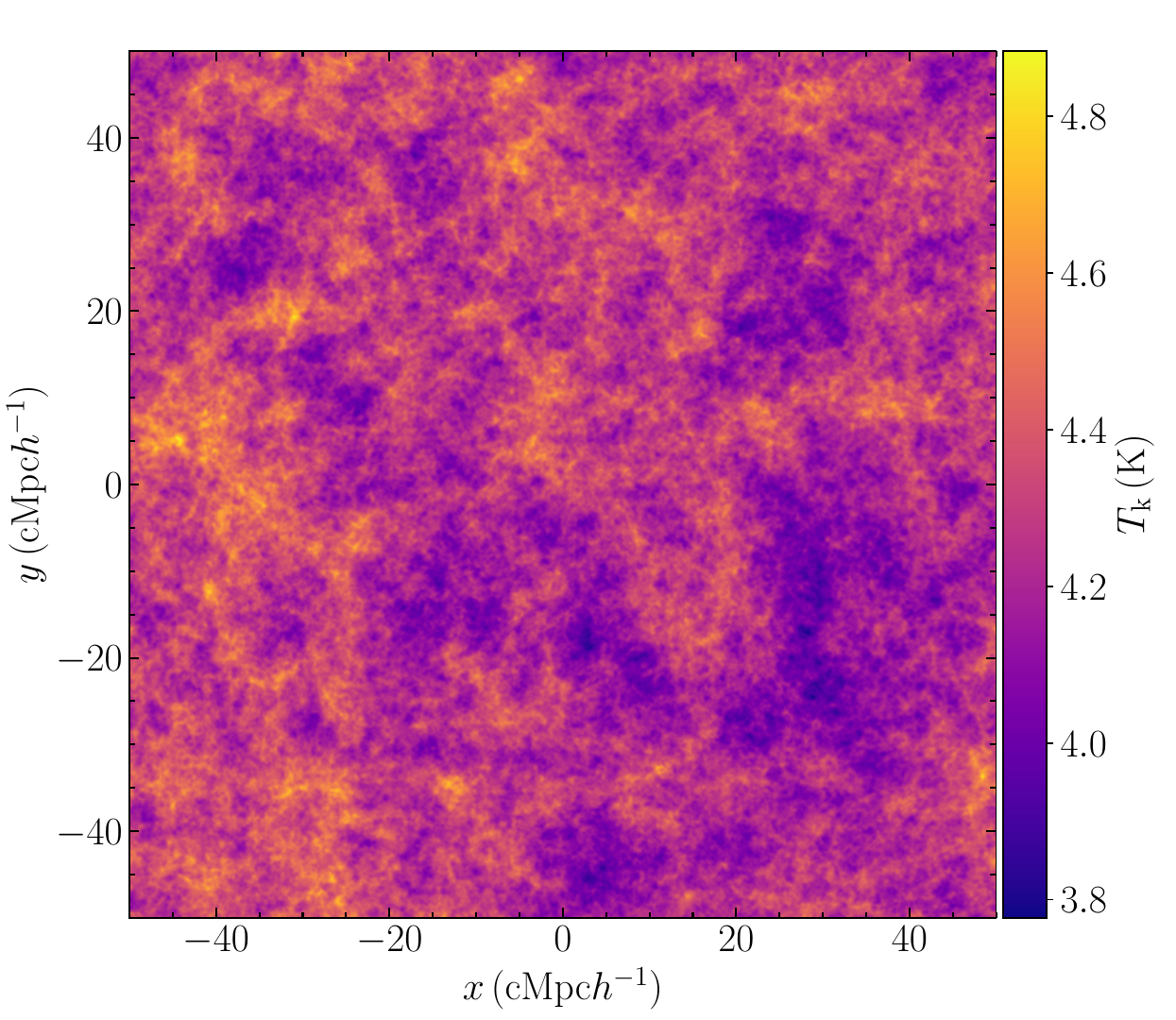}
\end{subfigure}
\caption{Left: projection-averaged gas overdensity, $1+\delta$ along the length in $\protect\cz$ axis of our \texttt{RAMSES} simulation box. The white circles mark dark matter haloes, which serve as sources of UV photons in our subsequent RT computations. Circle size corresponds to the mass of the halo. Right: projection-averaged gas kinetic temperature $T_{\mathrm{k}}\,$(K) along the length in $\protect\cz$ axis of the simulation box. The box is a unigrid cube at $z\approx13$ of length $100\,\mathrm{cMpc}h^{-1}$ on each side, with $1024^3$ cells.}\label{den}
\end{figure*}

\subsection{21-cm signal}
Readers new to 21-cm physics are referred to more detailed accounts in reviews such as those by \citet{Furlanetto} and \citet{Pritchard_2012}; here we give only a brief summary. The 21-cm signal is a measurement of the 21-cm brightness against the CMB. In our previous papers \citep{Mittal_2020, Mittal_2021, Mittal_2022b} we worked with the `global' signal, which is an average of the 21-cm brightness contrast over a cosmological volume. Here we work with a 21-cm signal at position $\mathbfit{r}$ and redshift $z$, which is approximately given by
\begin{multline}
T_{21}(\bm{r},z)=27x_{\ion{H}{i}}(\mathbfit{r},z)\left[1+\delta(\mathbfit{r},z)\right]\left(\frac{1-Y_{\mathrm{p}}}{0.76}\right)\left(\frac{\Omega_\mathrm{b}h^2}{0.023}\right)\\\times\sqrt{\frac{0.15}{\Omega_\mathrm{m}h^2}\frac{1+z}{10}}\left[1+\frac{1}{H(z)}\frac{\ud v_{\mathrm{b}}^{\parallel}}{\ud r^{\parallel}}\right]^{-1}\left[1-\frac{T_\gamma(z)}{T_\mathrm{s}(\mathbfit{r},z)}\right]\si{\milli\kelvin}\,,\label{DeltaT}
\end{multline} 
where $T_\mathrm{s}$ is the spin temperature, $T_\gamma$ is the CMB temperature, $x_{\ion{H}{i}}\equiv n_{\ion{H}{i}}/(n_{\ion{H}{i}}+n_{\ion{H}{ii}})$ is the ratio of number densities of neutral hydrogen (\ion{H}{i}) to the total hydrogen (H), and we have assumed a matter-dominated Universe so that $H(z)=H_0\sqrt{\Omega_\mathrm{m}(1+z)^3}$. The baryon overdensity is 1+$\delta(\mathbfit{r},z)=\rho_{\mathrm{b}}(\mathbfit{r},z)/\bar{\rho}_{\mathrm{b}}$, where $\bar{\rho}_{\mathrm{b}}$ is the mean of cosmic baryon density $\rho_{\mathrm{b}}(\mathbfit{r},z)$.

Gradient of the peculiar velocity (bulk velocity of the gas) along the line of sight is $\ud v_{\mathrm{b}}^{\parallel}/\ud r^{\parallel}$, where $v_{\mathrm{b}}^{\parallel}$ is the component of the bulk velocity along the line of sight and $r^{\parallel}$ is the proper distance along the line of sight \citep{Haimoud}. Note that $\ud v_{\mathrm{b}}^{\parallel}/\ud r^{\parallel}$ is a function of $\mathbfit{r}$ and $z$. In this work we choose our line of sight to be along the $\cz$ axis. Also, assuming the observer to be situated at an infinite distance we have
\begin{equation}
\frac{\ud v_{\mathrm{b}}^{\parallel}}{\ud r^{\parallel}}\approx\frac{\ud v^{\cz}_{\mathrm{b}}}{\ud \cz}\,,
\end{equation}
where $v^{\cz}_{\mathrm{b}}$ is the $\cz$ component of the bulk velocity at $\mathbfit{r}$ and $z$. For our fiducial setup (described below), the average, maximum and minimum value of the second to last term in equation~\eqref{DeltaT} are approximately $1.01$, $1.96$ and $0.80$, respectively.

In the presence of an excess radio background at frequencies $\sim\SI{1.4}{\giga\hertz}$ \citep{Fixsen_2011, Dowell_2018, Singal_2023} one should replace CMB with CMB plus excess radio background \citep{Feng_2018, Fialkov, Mittal_2022a}. However, in this work we do not include the contribution of such a radio intensity.

The spin temperature, $T_{\mathrm{s}}$, which quantifies the relative population of hyperfine levels in a neutral hydrogen atom, is given by
\begin{multline}
T_\mathrm{s}^{-1}(\mathbfit{r},z)\\=\frac{x_{\gamma}(\mathbfit{r},z) T_{\gamma}^{-1}(z)+[x_\mathrm{k}(\mathbfit{r},z)+x_\alpha(\mathbfit{r},z)]T_\mathrm{k}^{-1}(\mathbfit{r},z)}{x_{\gamma}(\mathbfit{r},z)+x_\mathrm{k}(\mathbfit{r},z)+x_\alpha(\mathbfit{r},z)}\,,\label{spin_temp}
\end{multline}
where $T_{\mathrm{k}}$ is the gas temperature, $x_{\gamma}$, $x_\mathrm{k}$ and $x_\alpha$ are the 21-cm, collisional and Ly$\alpha$ coupling, respectively. Because of a near thermal equilibrium of gas and Ly$\alpha$ photons, we have made an assumption that the colour temperature is equal to the gas kinetic temperature, i.e., $T_\alpha\approx T_{\mathrm{k}}$ \citep{Field}. We do not discuss 21-cm and collisional coupling any further, and point the interested reader to the details in our previous work \citep{Mittal_2021}.

Unlike in our previous work we have explicitly shown the position dependence on ionisation fraction, density and temperature (except for CMB, which is nearly uniform) in equations~\eqref{DeltaT} and \eqref{spin_temp} to emphasise that we are interested in a \emph{non-global} 21-cm signal. When we write $(\mathbfit{r},z)$ for any quantity (such as ionisation fraction, gas overdensity or gas temperature) derived from the simulated cosmological boxes, we mean the value of that field at the centre of a cell located at $\mathbfit{r}$ and redshift $z$ of the snapshot.

Besides the 21-cm signal itself we also look at its power spectrum. For 21-cm power spectrum we define the mean-subtracted 21-cm signal as (in units of temperature)
\begin{equation}
\delta_{21}(\mathbfit{r})=T_{21}(\mathbfit{r})-\left\langle T_{21}\right\rangle\,,
\end{equation}
where for brevity we dropped the $z$ dependence, and $\left\langle T_{21}\right\rangle$ represents the box or the space average of $T_{21}(\mathbfit{r})$. If the Fourier transform of $\delta_{21}(\mathbfit{r})$ is $\tilde{\delta}_{21}(\mathbfit{k})$ then the 21-cm power spectrum $P_{21}(k)$ is obtained via
\begin{equation}
\left\langle\tilde{\delta}_{21}(\bm{k_1})\tilde{\delta}_{21}(\bm{k_2})\right\rangle_{\mathrm{EA}}=(2\pi)^3\delta_{\mathrm{D}}(\bm{k_1}+\bm{k_2})P_{21}(\bm{k_1})\,,
\end{equation}
where $\delta_{\mathrm{D}}$ is the Dirac-delta function and $\langle\cdot\rangle_{\mathrm{EA}}$ represents the ensemble average. We will quantify only the isotropic fluctuations so that we focus on the spherically-averaged power spectrum in $k$ space, $P_{21}(\bm{k})= P_{21}(k)$. For a discrete system such as ours we compute $P_{21}(k)$ as
\begin{equation}
P_{21}(k_q)=(\Delta r)^3\left\langle|\tilde\delta_{21,q}^2|\right\rangle\,,
\end{equation}
where $\Delta r$ is the length of each cell and $\tilde\delta_{21,q}$ is the discrete Fourier transform of $\delta_{21}(\mathbfit{r})$ computed as
\begin{equation}
\tilde\delta_{21,q} = \frac{1}{n^{3/2}}\sum_{p=0}^{n-1}\delta_{21,p}\,\ue^{-\iota\bm{k}_q\cdot \bm{r}_p}\,,
\end{equation}
where $\delta_{21,p}=\delta_{21}(\mathbfit{r}_p)$, $r_p=p\Delta r$ and $k_q=2\pi q/(n\Delta r)$ for $q=0,1,2,\ldots ,(n-1)$ when there are $n$ divisions of each side (for our case this is 64).

Note that instead of $P_{21}$ we will illustrate our results in terms of
\begin{equation}
\Delta^2_{21}(k)=\frac{k^3}{2\pi^2}P_{21}(k)\,,
\end{equation}
which has the dimensions of temperature-squared.

\subsubsection*{Gas density, temperature and neutral hydrogen fraction}
We simulate the evolution and interaction of dark matter and gas via gravity, hydrodynamics and radiative cooling \& heating using the cosmological adaptive mesh refinement (AMR) code \texttt{RAMSES} \citep{Teyssier}. In \texttt{RAMSES} DM particles are collisionless particles that interact only via gravity and to track their evolution a collisionless Boltzmann solver is used. The dynamics of gas on the other hand is modelled using the hydrodynamical Euler equations (in their conservative form) coupled to DM through gravity. 

To get the number density of ionic species (only $n_{\ion{H}{i}}, n_{\ion{H}{ii}}$ and $n_{\mathrm{e}}$ are of interest to us) we use the default \texttt{RAMSES} thermochemistry solver where a collisional ionisation equilibrium -- but not thermal equilibrium -- is assumed. In this case, starting with a known primordial gas composition of hydrogen and helium, the number conservation along with the balance of creation and destruction rates uniquely determines all the abundances. The creation and destruction is governed by collisional ionisation and recombination, along with photoionisation if there is a non-zero ionising radiation present. Collisional ionisation, excitation, recombination and free-free emission (Bremsstrahlung) inevitably give rise to a gas cooling collectively known as radiative cooling. Besides these channels there is also the inverse Compton cooling. Photoionisation heating is active only in the presence of a non-zero ionising radiation. Note that all of the above mentioned processes are effective and important at high densities and for gas temperatures above $\SI{e4}{\kelvin}$ \citep{katz}. As we neither investigate high-density regions at small scales, nor do we model reionisation where temperatures are high, radiative cooling is unimportant and the gas temperature falls adiabatically in accordance with the Hubble expansion. This translates to $T_{\mathrm{k}}\propto(1+z)^2$, assuming an ideal gas law for the baryonic gaseous matter of adiabatic index $\gamma=5/3$ \citep{Scott}. In order to single out the effects of fully self-consistent 3D RT of Ly$\alpha$ photons we ignore X-ray heating or Ly$\alpha$ heating of IGM in this work.

We get our initial conditions using first-order Lagrangian perturbation theory with \citet{Eisenstein_1998} fit to the CDM transfer function with baryonic features at $z=99$.\footnote{We use the public initial-conditions-generator code \href{https://bitbucket.org/ohahn/monofonic/src/master/}{\texttt{monofonIC}} \citep{Michaux2020, hahn2020}.} We set our box size to $100\,\mathrm{cMpc}h^{-1}$ with $1024^3$ cells, which implies each cell being $97.6\,\mathrm{ckpc}h^{-1}$ in size. We have a total of $1024^3$ DM particles each of mass $6.88\times10^{7}\,\mathrm{M}_{\odot}h^{-1}$. As we are not interested in simulations of individual stars/galaxies or structure formation we work on a uni-grid system, so that there is no grid refinement, for all the results of this work\footnote{In the language of \texttt{RAMSES}, \verb|levelmin| and \verb|levelmax| are equal, so that no refinement takes place.}. After preparing initial conditions we run our hydrodynamic simulation down to $z\approx13$. Maps of gas overdensity and temperature are shown in Fig.~\ref{den}. We show a projected average of the quantity of interest which can be defined as
\begin{equation}
\tilde{f}(x,y)=\frac{\int f(x,y,\cz)\ud \cz}{\int \ud \cz}\,,
\end{equation}
if projection is done along the $\cz$ axis. The left panel shows the gas overdensity, $1+\delta$, the colour bar for which is in logarithmic scale. The white circles mark the DM haloes, found using HOP algorithm, as discussed below. The dark matter haloes will serve as sources of photons in our work (more on this in Section~\ref{source}). The average, maximum and minimum gas overdensity for this box are approximately $0.92, 78.6$ and $0.15$, respectively. The right panel shows the gas kinetic temperature, $T_{\mathrm{k}}$, the colour bar for which is in linear scale. The average, maximum and minimum temperature for this box are approximately $4.3, 38.8$ and $\SI{2.2}{\kelvin}$, respectively. Note that the mean temperature is consistent with the value expected from adiabatic evolution, i.e., $0.02(1+z)^2$ at $z\approx 13.2$ gives $\SI{3.8}{\kelvin}$. We use the python package \texttt{yt} \citep{Turk_2011} to process the \texttt{RAMSES} outputs.

In our work the DM haloes will serve as our sources for photons (more on this later). For low-resolution sizes, such as in this work, one may not find many collapsed objects at high redshifts. For this reason we do our analysis at a low redshift of $z\approx13$ for simulations. While in reality cosmic dawn is expected to be at higher redshifts, our set-up is sufficiently useful in model building and for drawing useful conclusions.

\subsubsection*{Ly\texorpdfstring{$\alpha$}{α} coupling}
We now discuss the Ly$\alpha$ coupling, $x_\alpha$. The probability that scattering of Ly$\alpha$ photon off a neutral hydrogen atom will cause a hyperfine transition is $4/27$ \citep{Meiksin00, Hirata, dijkstra_2008}. If the rate of scattering of Ly$\alpha$ photons per atom is $P_\alpha$ then the Ly$\alpha$ coupling term is \citep{Furlanetto}
\begin{equation}
x_\alpha=\frac{T_*}{T_\gamma}\frac{4P_\alpha}{27A_{10}}\,,\label{xalpha}
\end{equation}
where $T_{*}= \SI{0.068}{\kelvin}$ and $A_{10}=\SI{2.85e-15}{\second^{-1}}$ is the Einstein coefficient of spontaneous emission for the hyperfine transition. The scattering rate per atom may be calculated as
\begin{equation}
P_\alpha(\bm{r})=4\pi\int_{0}^{\infty} J(\bm{r},\nu)\sigma_{12}(\nu) \ud\nu\,,\label{pa}
\end{equation}
where $J$ and $\sigma_{12}$ are the local specific intensity (by number) and the cross-section of Ly$\alpha$ photons of frequency $\nu$, respectively. We discuss the numerical computation of $P_\alpha$ in greater detail in Section~\ref{sec:pa}.

\subsection{The sources of Lyman-series photons}\label{source}
This section gives the description of the sources and the luminosity $L$ of our Lyman-series photons. The fundamental idea is that Lyman-series photons are produced by star forming galaxies that started appearing at cosmic dawn. Given the spectral energy distribution (SED) of the star/galaxy and an assumption that the emission rate should roughly follow the star formation rate we can construct the local emissivity function. We used a similar idea in our previous work to compute the uniform and homogeneous version of emissivity \citep{Mittal_2020, Mittal_2021}. Our first ingredient required is the star formation rate density (SFRD).

Given the lack of full understanding of star and galaxy formation at cosmic dawn and the required computational expense we implement sources of radiation in a simplified fashion in our haloes. Once these DM haloes are located (as shown in the left panel of Fig.~\ref{den} by white circles), we assume that each halo contributes to the SFR of the full box in proportion to its mass. Following a simple analytical prescription for the SFRD \citep{F06}, the SFR due to $j^{\mathrm{th}}$ halo is
\begin{equation}
\psi_{j}=\frac{M_j}{M_{\mathrm{tot}}}\times V_{\mathrm{box}}f_{\star}\bar{\rho}_{\mathrm{b}}\frac{\ud F_{\mathrm{coll}}(z)}{\ud t}\,,\label{sfrd}
\end{equation}
where $f_{\star}$ is the star formation efficiency, $\bar{\rho}_{\mathrm{b}}$ is the mean cosmic baryon density today, $V_{\mathrm{box}}$ is the comoving box volume, $M_j$ is the mass of the $j^{\mathrm{th}}$ halo and $ M_{\mathrm{tot}}=\sum_j M_j$ is the sum of the mass of all the haloes found at the current snapshot.

We use a halo finder\footnote{We use the python package \texttt{yt-astro-analysis} for halo finding \citep[][\url{https://doi.org/10.5281/zenodo.5911048}]{Turk_2011}.} that implements the HOP algorithm \citep{Hut_1998} using a density threshold of $\delta_{\mathrm{peak}}=100$. For the snapshot shown in Fig.~\ref{den}, we find the minimum and maximum halo masses $2.89\times10^9\,\mathrm{M}_{\odot}h^{-1}$ and $2.20\times10^{10}\,\mathrm{M}_{\odot}h^{-1}$, respectively and a total of 34 haloes.


The star formation efficiency is a measure of the fraction of baryons that collapsed into the haloes and converted into star particles. In this work we take it to be $f_\star=0.1$ throughout. The fraction of DM that has collapsed into haloes is
\begin{equation}
F_{\mathrm{coll}}(z)=\mathrm{erfc}\left[\frac{\delta_{\mathrm{crit}}(z)}{\sqrt{2}\sigma(m_{\mathrm{min}})}\right]\,,    
\end{equation}
where $\delta_{\mathrm{crit}}$ is the linear critical overdensity of collapse and $\sigma^2$ is the variance in smoothed density field. The minimum halo mass for star formation is \citep{BL01}
\begin{equation}
m_{\mathrm{min}}=\frac{10^8\mathrm{M}_{\odot}}{\sqrt{\Omega_{\mathrm{m}}h^2}}\left[\frac{10}{1+z}\frac{0.6}{\mu}\frac{\mathrm{min}(T_{\text{vir}})}{\num{1.98e4}}\right]^{3/2}\,.
\end{equation}
For atomic cooling threshold min($T_{\text{vir}})=\SI{e4}{\kelvin}$ and a neutral medium at cosmic dawn $\mu\approx1.22$. We do not include any feedback effects, such as the Lyman--Werner feedback.

If there are $N_{\mathrm{haloes}}^i$ haloes in a cell $i$ (cell location $\bm{r}_i$), then the local SFRD at $\bm{r}_i$ is 
\begin{equation}
\dot{\rho_\star}(\bm{r}_i,z) = \frac{1}{V_\mathrm{cell}}\sum^{N_{\mathrm{haloes}}^i}_{j=1}\psi_j\,,
\end{equation}
where $V_\mathrm{cell}$ is the comoving volume of the $i^\mathrm{th}$ cell. With the above procedure we are finally able to obtain our SFRD in each cell and hence the local SFRD function, $\dot{\rho_\star}=\dot{\rho_\star}(\bm{r},z)$.

Next we need the rest-frame SED of Lyman-series photons emitted from the stars and galaxies. Our knowledge of properties of stars and star formation at cosmic dawn is only speculative. Stellar population synthesis data of, such as BPASS or \citet{bc03}, are calibrated to low redshift observations only. Thus, the employment of such data to high redshift is as good as any other model. We set SED of emission from any of these sources, residing in haloes, to be the same, i.e., independent of halo mass, metallicity or age. Ly$\alpha$ photons can be generated in two ways: either the photons emitted by the source between Ly$\alpha$ and Ly$\beta$ frequencies get redshifted to Ly$\alpha$ photons, which we know as continuum photons, or via radiative cascading of higher Lyman-series photons, which are known as injected photons \citep{Chen}. In this work we do not account for the injected photons, i.e., our sources do not emit any photons beyond Ly$\beta$\footnote{We do not have any additional Lyman-continuum background from faraway sources outside the box.}. Our rest-frame SED -- $\phi(\nu)$, defined in terms of number of photons per baryonic particle per unit frequency range -- follows a Pop-II model so that it is proportional to $\nu^{-0.86}$ extending from Ly$\alpha$ to Ly$\beta$ frequencies with the normalisation set to $N_{\alpha\beta}=6520$ photons per baryonic particle \citep{BL05}. As pointed out by \citet[][hereafter \citetalias{Zheng}]{Zheng} owing to the short range of continuum photons the results should only be mildly dependent on our SED choice. In the case with multiple scatterings we account for the velocity of the sources (which are DM haloes for us) so that frequency in the rest-frame of the source, $\nu_{\mathrm{source}}$, is linked to the frequency in global frame, $\nu$, by an appropriate Doppler factor as follows
\begin{equation}
\nu=\nu_{\mathrm{source}}\left(1+\frac{\mathbfit{v}_{\mathrm{source}}}{c}\cdot\hat{\mathbfit{k}}_{\mathrm{em}}\right)\,,\label{emitted_freq}
\end{equation}
where $\mathbfit{v}_{\mathrm{source}}$ is the source velocity, $c$ is the speed of light, and $\hat{\mathbfit{k}}_{\mathrm{em}}$ is the unit vector in the direction of emission of the photon. Note that $\hat{\mathbfit{k}}_{\mathrm{em}}$ is always drawn from an isotropic distribution. We have used a non-relativistic version of Doppler effect as $v_{\mathrm{source}}\ll c$. For our simulation the typical source velocities range from 30 to $\SI{300}{\kilo\metre\second^{-1}}$. 

Finally, with the SED and SFRD in place, the local emissivity (in terms of number of photons per unit comoving volume per unit time per unit frequency) at redshift $z$ and frequency $\nu$ is
\begin{equation}
\epsilon(\bm{r},\nu,z)=\frac{1}{m_{\mathrm{b}}}\dot{\rho}_{\star}(\bm{r},z)\phi(\nu)\,,\label{emiss}
\end{equation}
where $m_{\mathrm{b}}\approx1.22m_{\mathrm{H}}$ is the average baryon mass with $m_{\mathrm{H}}$ being the hydrogen mass. Consequently, the total luminosity in units of number of photons per unit time (required for equation~\ref{palpha}) in a box of comoving volume $V_{\mathrm{box}}$ can be written as
\begin{equation}
L=V_{\mathrm{box}}N_{\alpha\beta}f_{\star}\bar{n}_{\mathrm{b}}\frac{\ud F_{\mathrm{coll}}(z)}{\ud t}\,,\label{lum}
\end{equation}
where $\bar{n}_{\mathrm{b}}$ is the global average baryon comoving number density. As an example, at $z\sim13$ and for a box of side length $100\,\mathrm{cMpc}h^{-1}$ we have $L=\SI{3.6e58}{\second^{-1}}$.

\subsection{Radiative transfer of Ly\texorpdfstring{$\alpha$}{α} photons}\label{mcrt}
We use \texttt{RASCAS} \citep{rascas} to post-process our \texttt{RAMSES} simulation (thus accounting for cosmological density, temperature and bulk motion) to do a propagation of Ly$\alpha$ photons with multiple scatterings. We do a Ly$\alpha$ radiative transfer on our box smoothed by a factor of 16, so that we effectively have $64^3$ cells instead of $1024^3$.
  
We use the version presented by \citet{Garel21} that includes an implementation of the Hubble flow, but we make three modifications described in Sections \ref{sec:hf}, \ref{sec:sc} and \ref{sec:pa}, respectively -- (i) the introduction of a global comoving frame in which photons redshift during their propagation, (ii) stopping criterion for photon propagation, and (iii) the computation of the scattering rate per atom.

We give here a general description of the algorithm of Monte Carlo radiative transfer (MCRT) of Ly$\alpha$ photons. Readers familiar with the RT details may skip and jump straight to the results in Section~\ref{results}. Before we get to the more detailed version of MCRT algorithm we describe three types of reference frames of interest. These are as follows:
\begin{itemize}
\item Global comoving frame: an observer in this frame can see both contributions to the velocity, viz., macroscopic bulk or peculiar velocity $(\mathbfit{v}_{\mathrm{b}})$ and microscopic thermal motion $(\mathbfit{v}_{\mathrm{th}})$ of the hydrogen atom. One can think of this observer to be sitting at the corner of the cosmological box observing all the events happening. We put no subscript on the photon's frequency in this frame.
\item Cell or gas frame: an observer in this frame cannot see the bulk velocity of gas but only the thermal velocity. It is in this frame that the line profile of atomic transition is a Voigt function. We use the subscript `cell' on $\nu$ to represent the frequency of photons in this frame. Thus,
\begin{equation}
\nu_{\mathrm{cell}}=\nu\left(1-\frac{\mathbfit{v}_{\mathrm{b}}}{c}\cdot\hat{\mathbfit{k}}\right)\,,
\end{equation}
where $\hat{\mathbfit{k}}$ is the photon's direction of propagation. In the above equation -- and for similar ones that follow -- we used the linearised version of the Lorentz transformation as appropriate for non-relativistic velocities.
\item Atom frame: as the name suggests, the observer does not see any velocity and the atom is at rest. The line profile in this frame is simply the natural (Lorentzian) line profile. We use the subscript `atom' on $\nu$ to represent the frequency in this frame. Thus,
\begin{equation}
\nu_{\mathrm{atom}}=\nu_{\mathrm{cell}}\left(1-\frac{\mathbfit{v}_{\mathrm{th}}}{c}\cdot\hat{\mathbfit{k}}\right)=\nu\left(1-\frac{\mathbfit{v}_{\mathrm{b}}+\mathbfit{v}_{\mathrm{th}}}{c}\cdot\hat{\mathbfit{k}}\right)\,,
\end{equation}
where we neglected the second-order term in velocity.
\end{itemize}

We now discuss the main steps in MCRT describing the important concepts along the way.
\begin{enumerate}
\item[1)] \textit{Initialising the photon:} a photon is started from the source of known position. We assign it a direction $(\hat{\mathbfit{k}}_{\mathrm{em}})$ chosen from an isotropic distribution and a frequency (in the source frame) chosen from a given SED of the source, i.e., spectral sampling from the SED \citep{rascas}. Then we translate the frequency to external frame according to the source velocity. See equation~\eqref{emitted_freq}.

\item[2)] \textit{Propagating the photon:} after emission (or a scattering event) we assign the photon a new optical depth, which we choose from an exponential distribution, i.e., $\tau_{\mathrm{scat}}=-\ln(1-r)$, for a random number $r\in[0,1)$. We then move the photon a real physical distance along its current direction of propagation such that the optical depth it covers is $\tau_{\mathrm{scat}}$. At the new location, a resonant scattering occurs.

\citet{rascas} moves the photons from event to event depending on the ambient density, temperature, and the photon frequency until the optical depth accumulates to $\tau_{\mathrm{scat}}$. By an `event' we mean either a cell-crossing event or a scattering event. For a distance $l$ inside a cell of number density $n_{\ion{H}{i}}$ and gas temperature $T_{\mathrm{k}}$ the optical depth for the Ly$\alpha$ photon is
\begin{equation}
\tau=n_{\ion{H}{i}}\sigma_{12}(\nu,T_{\mathrm{k}})l\,.\label{od1}
\end{equation}
In the above we assumed that all hydrogen atoms are in the ground state, so that $n_{\ion{H}{i},1}\approx n_{\ion{H}{i}}$, because of the smallness of transition time ($A_{21}^{-1}\sim\SI{e-9}{\second}$) compared to other times scales and the high characteristic temperature for $1\to2$ excitation compared to $T_{\mathrm{k}}$. Note that we do not assume any deuterium or dust in the IGM gas in any of the models in our work.

The photon propagation picture presented above is applicable to a non-Hubble-expanding system. With Hubble flow in action the optical depth calculation needs modification and is discussed in Section~\ref{sec:hf}.

Regardless of the inclusion of Hubble flow, in MCRT codes one finds it easiest to work in gas/cell frame when computing the optical depth, in which case the cross-section is represented by a Voigt profile, i.e., the convolution of natural (Lorentzian) and thermal (Gaussian) line profiles. Thus,
\begin{equation}
\sigma_{\mathrm{V}}(\nu_{\mathrm{cell}},T_{\mathrm{k}})=\frac{3\lambda_{\alpha}^2a}{2\sqrt{\pi}}H_a(x_{\mathrm{cell}})\,,
\end{equation}
where $x_{\mathrm{cell}}=(\nu_{\mathrm{cell}}-\nu_\alpha)/\Delta\nu_\mathrm{D}$ and $a=A_{21}/(4\pi\Delta\nu_\mathrm{D})$ is the ratio of natural to thermal line broadening for $A_{21}=\SI{6.25e8}{\second^{-1}}$, the Einstein spontaneous emission coefficient of Ly$\alpha$ transition. The line centre frequency for Ly$\alpha$ is $\nu_\alpha=\SI{2.47e15}{\hertz}$. The Voigt function is defined as
\begin{equation}
H_a(x)=\frac{a}{\pi}\int_{-\infty}^{\infty}\frac{\ue^{-y^2}}{(x-y)^2+a^2}\ud y\,,\label{voigt}
\end{equation}
which is a non-dimensional function normalised such that $\int H_a(x)\ud x=\sqrt{\pi}$. The temperature dependence enters through the Doppler width $\Delta\nu_\mathrm{D}=b/\lambda_\alpha$, where $b=(2k_{\mathrm{B}}T_{\mathrm{k}}/m_{\mathrm{H}})^{1/2}$ is the mean thermal velocity for hydrogen atom of mass $m_{\mathrm{H}}$ and $\lambda_\alpha=\SI{1215.67}{\angstrom}$ is the central wavelength.

\item[3)] \textit{Scattering:} after we have moved the photon by a physical distance governed by $\tau_{\mathrm{scat}}$ a scattering event occurs where we assign the photon a new direction and frequency. The angle by which the photon is scattered is decided by the phase function. If $\mu=\cos\theta$, where $\theta$ is the scattering angle or the angle between the outgoing and incoming direction of the photon, then the probability of $\cos\theta$ to be in $\mu$ to $\mu+\ud\mu$ is $\mathcal{P}(\mu)\ud\mu$. For the Ly$\alpha$ line, there are two limiting cases depending on whether the photon is in the core or wings of the line profile as seen by the atom just before scattering \citep{Dijkstra08},
\begin{equation}
\mathcal{P}(\mu)=
\begin{cases}
(11+3\mu^2)/24,\text{ if } |\nu_{\mathrm{atom,in}}-\nu_\alpha|<0.2\Delta\nu_{\mathrm{D}}\\
3(1+\mu^2)/8,\text{ otherwise}\,,
\end{cases}\label{phase}
\end{equation}
and $\mu$ goes from $-1$ to 1. For simplicity, we label this as anisotropic scattering. Note that under the assumption of non-relativistic velocities the angle of scattering $\theta$ is the same in any frame. 

Without the loss of generality, let us align the incoming photon along the positive $x$ axis and the outgoing photon in $xy$ plane, so that $\hat{\mathbfit{k}}_{\mathrm{in}}=(1,0,0)$ and $\hat{\mathbfit{k}}_{\mathrm{out}}=(\mu,\sqrt{1-\mu^2},0)$. In this coordinate system only the $x$ component (parallel) and $y$ component (perpendicular) of thermal velocity are important for the determination of $\hat{\mathbfit{k}}_{\mathrm{out}}$.

The choice of parallel velocity component will depend on the frequency of the incoming photon. If the incoming photon is close to line centre, photon will favour an atom with a velocity such that the photon appears closer to the line centre, where the interaction cross-section is high. On the other hand if the incoming photon has a large frequency offset it is more likely to get scattered by a slow moving atom. This is because faster atoms will be rarer to find which can satisfy the large frequency offset. Accordingly, the probability distribution of parallel component is given by a 1D Gaussian convolved by an appropriate Lorentzian as follows \citep{Dijkstra_2006, Laursen_2009}
\begin{equation}
P_{\parallel}(u^{\parallel}_{\mathrm{th}})=\frac{1}{H_a(x_{\mathrm{cell,in}})}\frac{a}{\pi}\frac{\ue^{-{u^{\parallel}_{\mathrm{th}}}^2}}{\left(x_{\mathrm{cell,in}}-u^{\parallel}_{\mathrm{th}}\right)^2+a^2}\,,\label{parallel}
\end{equation}
where $u^{\parallel}_{\mathrm{th}}=v^{\parallel}_{\mathrm{th}}/b$ is the non-dimensionalised parallel component of thermal velocity. 

The perpendicular component, on the other hand, is not seen by the photon and hence we choose it from a pure 1D Maxwell--Boltzmann distribution, which is simply a Gaussian function. Thus,
\begin{equation}
P_{\perp}(u^{\perp}_{\mathrm{th}})=\frac{1}{\sqrt{\pi}}\ue^{-{u^{\perp}_{\mathrm{th}}}^2}\,.\label{perpendicular}
\end{equation}

Having obtained $\mu$ (and hence $\hat{\mathbfit{k}}_{\mathrm{out}}$) and $\mathbfit{v}_{\mathrm{th}}=(bu^{\parallel}_{\mathrm{th}},bu^{\perp}_{\mathrm{th}},\text{`not important'})$ the new frequency to first order in $\mathbfit{v}_{\mathrm{th}}/c$ is 
\begin{equation}
\nu_{\mathrm{cell,out}}=\nu_{\mathrm{cell,in}}\frac{1+(\hat{\mathbfit{k}}_{\mathrm{out}}-\hat{\mathbfit{k}}_{\mathrm{in}})\cdot\mathbfit{v}_{\mathrm{th}}/c}{1+(1-\hat{\mathbfit{k}}_{\mathrm{out}}\cdot\hat{\mathbfit{k}}_{\mathrm{in}})h_{\mathrm{P}}\nu_{\mathrm{cell,in}}/m_{\mathrm{H}}c^2}\,,\label{nuout}
\end{equation}
where $h_{\mathrm{P}}$ is the Planck's constant. Numerator is just the term due to a change of frames assuming a coherent scattering in atom frame, i.e., $\nu_{\mathrm{atom,in}}=\nu_{\mathrm{atom,out}}$. But if recoil in the atom is taken into account then there is a partial transfer of energy from photon to atom, which means the outgoing frequency is slightly lowered by the factor in the denominator \citep{adams1972, Zheng_2002}. As we discuss below Figure~\ref{fig:rascstest1} illustrates the effect of recoil on the output specific intensity in an idealised configuration.

\item[4)] \textit{Stopping criterion:} we repeat steps 2 and 3 until the IGM becomes transparent enough for the photon. We define the transparency quantitatively in Section~\ref{sec:sc}.
\end{enumerate}

We repeat all steps for a desired number of Monte Carlo (MC) photons.

Some general remarks about our RT approach are in order. \texttt{RASCAS} is a passive RT code, i.e., RT does not have any effect on the hydrodynamics or in other words the gas density and temperature distribution provided by \texttt{RAMSES} serve as a fixed background on which RT is run.

Related to the above point, note that when we write `recoil in the atom' it is mentioned here only as a theoretical concept meant to explain the decrease in photon energy. In actual computation in the code, the energy transfer is a one-sided process which affects only the photon since \texttt{RASCAS} is a passive RT code run in post-process. If indeed the energy gain in the atom is taken into account then it gives rise to the so-called `Ly$\alpha$ heating'. However, this heating is quite small and we do not expect the energy conservation violation to be severe.

A final important point we mention is that, contrary to the typical use cases of \texttt{RASCAS} code, our purpose of RT simulation is not to calculate the Ly$\alpha$ spectrum escaping some domain but rather the rate of scattering, $P_\alpha$, over the full domain volume.

\begin{figure}
\centering
\begin{tikzpicture}[scale=2]
\draw (-2,0) -- (-1,0) -- (-1,1) -- (-2,1) -- (-2,0);
\draw (-1,0) -- (0,0) -- (0,1) -- (-1,1);
\node[above] at (-1.5,1) {$c_1$};
\node[above] at (-0.5,1) {$c_2$};
\begin{scope}[thick,decoration={markings,mark=at position 0.5 with {\arrow{stealth}}}] 
\draw[postaction={decorate}] (-1.5,0.5)-- (-0.76,0.8); 
\draw[postaction={decorate}] (-0.76,0.8) -- (-0.61,0.48);
\draw[postaction={decorate}] (-0.61,0.48) -- (0.39,0.6);
\draw[postaction={decorate}] (-1.5,0.5) -- (-0.3,-0.1);
\end{scope}
\node[above] at (-1.5,0.5) {0};
\node[left] at (-1,0.75) {1};
\node[above] at (-0.76,0.8) {2};
\node[below] at (-0.61,0.48) {3};
\node[left] at (0,0.65) {4};
\node[below left] at (-1,0.25) {5};
\node[above] at (-0.5,0) {6};
\draw [fill,blue] (-1,0.702) circle [radius=0.025];
\draw [fill,red] (-0.76,0.8) circle [radius=0.025];
\draw [fill,red] (-0.61,0.48) circle [radius=0.025];
\draw [fill,blue] (0,0.5532) circle [radius=0.025];
\draw [fill,blue] (-1,0.25) circle [radius=0.025];
\draw [fill,blue] (-0.5,0) circle [radius=0.025];
\end{tikzpicture}
\caption{This schematic demonstrates our calculation of the rate of scattering per atom, $P_{\alpha}$. Shown are trajectories of two example photons in cells $c_1$ and $c_2$ from our cosmological box. Red and blue dots mark scattering and cell-crossing events, respectively. Both photons originate from location `0'. (Although this schematic is shown for 2D, it is easy to visualise the concept in 3D.)}\label{p_scheme}
\end{figure}

\subsubsection{Hubble flow}\label{sec:hf}
\citet{rascas} did not account for redshifting of photons. Often in MCRT codes one works in global rest frame to capture the effect of Hubble expansion. In this case scatterers are given an additional Hubble flow velocity keeping the frequency of the photon unchanged as it freely propagates between scatterings. This is a reasonable approach as long as the mean free path of the photons is small. We instead work in a comoving frame when the photon is in free propagation. In such a case the observer does not see any additional Hubble flow velocity of atoms but directly redshifts the photons as they propagate.

With redshifting in place an additional complication arises in either of the approaches mentioned above: redshifting the photons corresponding to a large distance $l$ in just one step may cause large shifts in $x_{\mathrm{cell}}$ even though $\nu$ may change by a very small amount. This may cause the photon to miss the essential core scatterings where the line profile is sharply peaked. Hence, following \citet{Garel21}, we use an adaptive scheme to accurately account for the \ion{H}{i}-Ly$\alpha$ overlap. In this method, instead of moving photons over large steps of length $l$ we move them over sub-cell lengths, $\delta l$. For sufficiently small $\delta l$ one correctly captures the redshifting by altering the frequency by a factor of $(1-H\delta l/c)$ without missing the essential core scatterings.

Consider a photon at redshift $z_0$ of frequency $\nu_0$ at some location and after a short time interval $\delta t$ it reaches $z_1$ at frequency $\nu_1$. In a Hubble expanding medium, the frequencies can be related as
\begin{equation}
\frac{\nu_{1}}{1+z_1}=\frac{\nu_0}{1+z_0}\,.
\end{equation}
For a small distance travelled by the photon, $\delta l$, the change in redshift is $z_1-z_0=\delta z$, where $|\delta z|\ll1$. We get
\begin{equation}
\nu_1=\nu_0\left(1+\frac{\delta z}{1+z_0}\right)\,.
\end{equation}
Eliminating the $z$'s using the definition of Hubble factor and writing in terms of $\delta t$, we get $\nu_1=\nu_0(1-H\delta t)$. Finally, setting $\delta t=\delta l/c$ the above becomes
\begin{equation}
\nu_1=\nu_0\left(1-\frac{H\delta l}{c}\right)\,.\label{redshift}
\end{equation}

With this new scheme of propagation equation~\eqref{od1} is modified to
\begin{equation}
\tau=\sum_i n_{\ion{H}{i}}\sigma_{\mathrm{V}}(\nu_i,T_{\mathrm{k}})\delta l_i\,,
\end{equation}
where $\nu_{i+1}$ is related to previous frequency $\nu_{i}$ via equation~\eqref{redshift}.

For a quantitative comparison of the two implementations -- first in which the gas atoms are given Hubble flow velocity and the second in which we directly reduce the photon frequency -- see Section~\ref{laursen}.

\subsubsection{IGM transparency}\label{sec:sc}
As we use periodic boundary conditions for the photons, we give details of the stopping criterion of our MCRT procedure. We terminate the MCRT in step 4 when the IGM becomes transparent enough or in other words when the photon has drifted far to the red wings of the Voigt line profile. Quantitatively we set the stopping criterion to be when the cell frame frequency, $\nu_{\mathrm{cell}}$, goes below $(\nu_\alpha-8\Delta\nu_{\mathrm{D}})$, i.e., when $x_{\mathrm{cell}}\leqslant-8$. In such a case the cross-section at the critical frequency relative to that at the line centre is, for example, $\sigma_{\mathrm{V}}(\nu_{\mathrm{cell}},T_{\mathrm{k}})/\sigma_{\mathrm{V}}(\nu_{\alpha},T_{\mathrm{k}})\approx 4.5\times10^{-4}$ and $4.3\times10^{-5}$ at $T_{\mathrm{k}}=1$ and $\SI{100}{\kelvin}$, respectively.

\subsubsection{Rate of Ly\texorpdfstring{$\alpha$}{α} scattering}\label{sec:pa}
In numerical simulations it can be computationally expensive to record $J(\bm{r},\nu)$ because it is a four dimensional quantity. It is thus simpler to work directly with $P_\alpha(\bm{r})$ in order to get $x_{\alpha}$. Following \citet{Lucy99} and \citet{Seon_2020}, we compute $P_\alpha$ in a cell by summing optical depth segments due to all the MC photons that visited that cell times the luminosity (by number) carried by each MC photon normalised to the number of atoms in that cell. Thus, the scattering rate per atom in a cell `X' is
\begin{equation}
P_\alpha(\mathrm{X})=\frac{L/N_{\mathrm{mc}}}{V_{\mathrm{X}}n_{\ion{H}{i},\mathrm{X}}}\sum\tau_{\mathrm{X}}\,,\label{palpha}
\end{equation}
where $L$ is the total box luminosity by number (equation~\ref{lum}), $N_\mathrm{mc}$ is the number of MC photons used, $V_{\mathrm{X}}$ is the volume of cell X, $n_{\ion{H}{i},\mathrm{X}}$ is the neutral hydrogen number density in cell X and $\sum\tau_\mathrm{X}$ is the total optical depth in cell X due to all the MC photons that crossed this cell. Note that a slightly more accurate approach to $P_\alpha$ is to count the number of scatterings rather than summing the optical depths in which case one must replace $\sum\tau_{\mathrm{X}}$ by $N_{\mathrm{scatt,X}}$, the number of scatterings in each cell \citep{Seon_2020}. However, we work with the definition in equation~\eqref{palpha} for reasons discussed in appendix~\ref{pointvspath}.

Figure~\ref{p_scheme} illustrates the scheme of equation~\eqref{palpha} diagrammatically in a toy model consisting of 2 cells $c_1$ and $c_2$ and 2 MC photons. The cell crossings are marked by blue dots and scattering events are shown by red dots. In cell $c_1$ there are 2 segments of the photon trajectory namely $0\to1$ and $0\to5$. Accordingly the scattering rate following equation~\eqref{palpha} will be $$\frac{L/2}{V_{c_1}n_{\ion{H}{i},c_1}}(\tau_{01}+\tau_{05})\,.$$Similarly in cell $c_2$ there are 4 segments $1\to2, 2\to3, 3\to4$ and $5\to6$ and hence $$\frac{L/2}{V_{c_2}n_{\ion{H}{i},c_2}}(\tau_{12}+\tau_{23}+\tau_{34}+\tau_{56})\,,$$where $\tau_{ij}$ refers to the optical depth corresponding to the length segment $l_{ij}$ and $i,j$ label the scattering or cell-crossing event.

For convergence, one requires a large number of MC photons for a given number of cells. For our main results in this work, we have used $N_\mathrm{mc}=10^8$ and $64^3$ cells in a cube of side length $100\,\mathrm{cMpc}h^{-1}$. In Section~\ref{caveats} we discuss the MC noise associated with our RT simulations.

\subsection{Ly\texorpdfstring{$\alpha$}{α} coupling without multiple scatterings}\label{fft}
In Section~\ref{results}, we intend to compare the method presented in this paper (accounting for multiple Ly$\alpha$ scatterings) with our previous work \citep{Mittal_2020} which did not include this effect. Therefore, we briefly summarise the framework introduced in \citet{Mittal_2020} modified to 3D calculation below.

We use the following expression for the Ly$\alpha$ coupling term
\begin{equation}
x_\alpha=S_\alpha\frac{J_\alpha}{J_0}\,,\label{old_xa}
\end{equation}
where
\begin{equation}
J_0\approx\num{5.54e-8}(1+z)\,\si{\per\metre\squared\per\second\per\hertz\per\steradian}\,.
\end{equation}
In case of no multiple scatterings, the effect of line core scatterings to the Ly$\alpha$ coupling may be captured by the spectral distortion term, represented in literature as $S_\alpha$. \citet{Hirata} laid down a scheme to compute this factor accurately used in previous works by \citet{vonlanthen} and \citet{Semelin2023} who employ core-skipping algorithms in their radiative transfer simulations. However, analytical expressions exist in literature which approximate this factor reasonably, as has been verified by \citet{FP06}. We use the approximate form given by \citet{CS06}, so that 
\begin{equation}
S_\alpha=\exp(-1.69\zeta^{2/3})\,,
\end{equation}
where $\zeta=\sqrt{16\eta^3a\tau_\alpha/(9\pi)}$ is a function of redshift and local gas temperature. The temperature and redshift dependence enter through the Voigt parameter
$$a=a(T_{\mathrm{k}})\,,$$
the recoil parameter
$$\eta=\eta(T_{\mathrm{k}})\,,$$
and the Gunn--Peterson optical depth to Ly$\alpha$ resonance scattering
$$\tau_\alpha=\tau_\alpha(z)\,.$$
For details see equations~(13), (15) and (16) by \citet{Mittal_2020}.

In this formalism we need the background Ly$\alpha$ specific intensity, $J_\alpha(\bm{r},z)=J(\bm{r},z,\nu=\nu_\alpha)$, throughout the simulation box. In our previous works we used the following equation \citep{Mittal_2020} 
\begin{equation}
J_\alpha(z)=\frac{c}{4\pi}(1+z)^2\int\frac{\epsilon(\nu',z')}{H(z')}\,\ud z'\,,\label{j_old}
\end{equation}
for the computation of $J_\alpha$. However, the above is applicable only when we have a continuous uniform emission over the whole volume of interest; position information was irrelevant for the resulting $J_\alpha$. In our current system emission is not uniformly distributed over the volume. Instead, the emission occurs from specific sites (which in this work we set at the DM halo positions).

For 3D version of equation~\eqref{j_old}, we follow \citet{Santos_2008, santos2010} to get the spherically-averaged number of photons arriving per unit proper area per unit proper time per unit frequency per unit solid angle, 
\begin{equation}
J_\alpha(\bm{r},z)=\frac{1}{4\pi}(1+z)^2\int_0^{r_{\mathrm{max}}}\int\epsilon(\bm{r}+\bm{r}',\nu',z')\,\frac{\ud\Omega'}{4\pi}\ud r'\,,\label{J}
\end{equation}
where
\begin{equation}
\nu'=\nu_\alpha(1+z')/(1+z)\,,\label{nuprime}
\end{equation}
$z$ is the redshift of our snapshot and $z'$ is the redshift at the source from where the photons start and reach the point of interest $(\bm{r},z)$ while travelling in a straight line a comoving distance of $r'$. For a given $z$ and $r'$, $z'$ is obtained via
\begin{equation}
r'=\int_z^{z'}\frac{c}{H(z'')}\ud z''\,.\label{comoving}
\end{equation}
Note how $J_\alpha$ is independent of local gas density, local gas temperature and line profile shape -- and hence any off-centre scatterings -- since there is no scattering factor $\ue^{-\tau}$ in the computation of equation~\eqref{J}. The assumption $\tau=0$ originates from the notion of employing a Dirac-delta cross-section centred at resonant frequency. In such a scenario bluer photons have no overlap with the scatterer allowing the photons to travel long distances without getting scattered at all. They are scattered only when they cosmologically redshift to line centre.

Because $J_\alpha$ is independent of local physical quantities, $x_\alpha$ is also independent of them except weakly through $S_\alpha$, which for our current snapshot is $\sim0.67$. This formalism has been used in popular 21-cm codes such as \texttt{21cmFAST} \citep{Mesinger_11}.

For consistency with the SED used in this work we have not added the contribution of higher Lyman-series photons to the total intensity in equation~\eqref{J}. The maximum comoving distance, away from the desired position a photon could have started so that it reaches at Ly$\alpha$ frequency, allowed is $r_{\mathrm{max}}$. It may be computed by setting $\nu'$ to $\nu_\beta$ -- which is the frequency of Ly$\beta$ line -- in equation~\eqref{nuprime} and putting the corresponding $z'$ into equation~\eqref{comoving}.

\begin{figure*}
\centering
\includegraphics[width=0.9\linewidth]{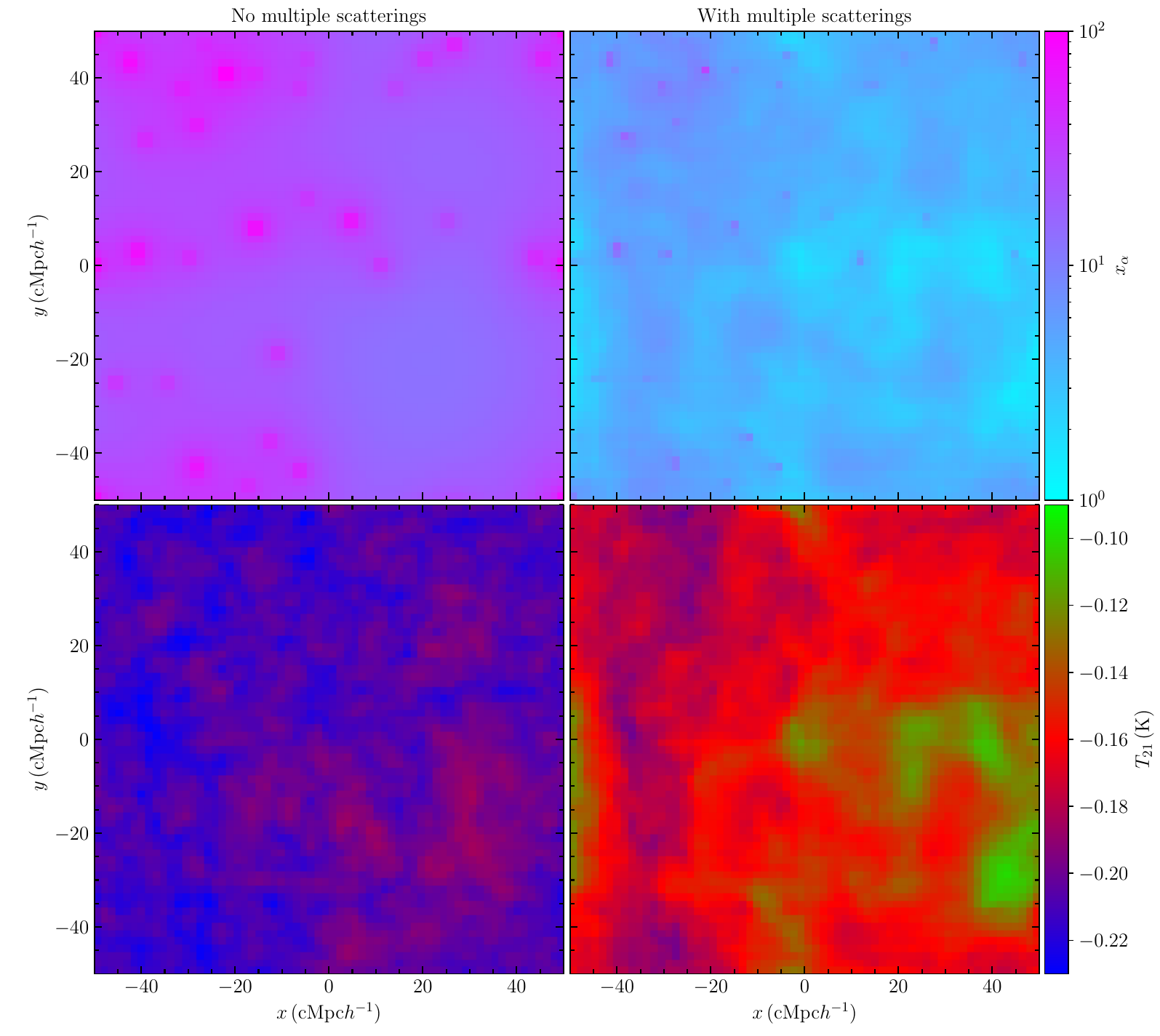}
\caption{Top panels show projected average along the full depth in $\protect\cz$ axis of the Ly$\alpha$ coupling $x_\alpha$, calculated without (left) and with (right) multiple scatterings. Bottom panels show the corresponding 21-cm signal $T_{21}$, projected average along the $\protect\cz$ axis. The box-averaged value of the 21-cm signal for left- and right-hand sides are $\langle T_{21}\rangle=\SI{-207.3}{\milli\kelvin}$ and $\SI{-159.2}{\milli\kelvin}$, respectively. The box is of size $100\,\mathrm{cMpc}h^{-1}$ with $64^3$ cells. The redshift of simulation is $z\sim13$.}\label{new_results}
\end{figure*}

Equation~\eqref{J} maybe more useful when written in terms of a single volume integral, emissivity in terms of SED-SFRD and with replacement $\bm{r}+\bm{r}' \to \bm{r}'$. Thus,
\begin{equation}
J_\alpha(\bm{r},z)=\frac{1}{4\pi}(1+z)^2\frac{1}{m_{\mathrm{b}}}\int \left[\dot{\rho}_\star(\bm{r}', z')\right]\left[\frac{\phi(\nu')}{4\pi |\bm{r}-\bm{r}'|^2}\right]\ud^3 r'\,.\label{Ja_conv}
\end{equation}
where $\nu'=\nu'(|\bm{r}-\bm{r}'|)$ and as before $z'$ is the higher redshift (obtained via equation~\ref{comoving} but with the replacement $r'\to|\bm{r}-\bm{r}'|$) corresponding to the epoch when the photons were emitted from star forming sites. The latter effect accounts for the delayed SFR. 
\begin{table*}
\centering
\caption{Simulations A to D designed to investigate the relative importance of different ingredients of Ly$\alpha$ radiative transfer. The bottom 5 rows list the values of $x_\alpha$ and $T_{21}$ in all simulations, and their difference from the fiducial simulation.}\label{RTeffects}
\def\arraystretch{1.1}
\begin{tabular}{l|rrrrr}
\hline
Process   & A & B & C & D & Fiducial \\ \hline
Bulk motion &  No & Yes & Yes & Yes & Yes\\
Cosmological \ion{H}{i} density &  Yes & No & Yes & Yes & Yes\\
Recoil  &  Yes & Yes & No & Yes & Yes\\
Anisotropic scattering &  Yes & Yes & Yes & No & Yes\\ \hline
Results\\ \hline
Mean difference from fiducial in $x_{\alpha}$  &  $4.86$ & $-0.38$ & $0.57$ & $0.01$ & 0\\
RMS difference from fiducial in $x_{\alpha}$ &  $7.47$ & $2.65$ & $0.90$ & $0.49$ & 0\\[1.5mm]
Box-averaged $T_{21}\,$(mK) &  $-194.1$ & $-156.9$ & $-163.7$ & $-159.7$ & $-159.2$\\
Mean difference from fiducial in $|T_{21}|\,$(mK) & $34.8$ & $-2.3$ & $4.5$ & $0.4$ & 0\\
RMS difference from fiducial in $|T_{21}|\,$(mK)  & $49.2$ & $8.7$ & $5.5$ & $4.4$ & 0\\ \hline
\end{tabular}
\end{table*}
The fact that the integral in equation~\eqref{Ja_conv} is a convolution of two functions, viz., $\dot{\rho}_\star$ and $\phi/(4\pi r^2)$, saves computation time, as one can use the convolution theorem of Fourier transforms to evaluate it.

\section{Results}\label{results}
Figure~\ref{new_results} shows our main result: the Ly$\alpha$ coupling (top panels) and the corresponding 21-cm signal (bottom panels). The left panels show the quantities without multiple scatterings (using the scheme in Section~\ref{fft}) and the right panels show them with multiple scatterings (using our RT simulation set-up described in Section~\ref{mcrt}).

Comparing the Ly$\alpha$ coupling maps visually one can immediately notice the differences. In the case without multiple scatterings, the photons have a vanishing cross-section away from the centre. Consequently, they have the freedom to travel large distances without getting scattered. So the photons travel in straight lines and can get scattered only when they reach the line centre beyond which they do not make any impact. In this case the gas density and temperature inhomogeneities are not reflected on $x_\alpha$ ($S_\alpha$ is only a very weak function of gas temperature) and consequently $x_\alpha$ has a smooth distribution away from the sources. On the other hand when there is a finite spread in the cross-section around the line centre photons undergo scatterings not only on the blue side but on the red side as well and continue to do so until the medium becomes transparent. As a consequence $x_\alpha$ picks up the density and temperature inhomogeneities as well as exhibits a non-spherical symmetry around the sources. Also, note that in the case with multiple scatterings the spin temperature is \textit{not} saturated to gas temperature and thus, fluctuations seen in the 21-cm signal on the right hand side are dominated by fluctuations in $x_\alpha$. This is not the case with no multiple scatterings where $x_\alpha$ on the overall is quite high because of which $T_{\mathrm{s}}\approx T_{\mathrm{k}}$; 21-cm signal fluctuations track the fluctuations in gas temperature.

The mean and root mean square (RMS) difference in $x_\alpha$ between the computation with multiple scatterings and that without is $\langle\Delta x_\alpha\rangle=19$ and $\sqrt{\langle(\Delta x_\alpha)^2\rangle}=30$, respectively. ($\Delta x_\alpha$ represents $x_{\alpha,\mathrm{No\ MS}} -x_{\alpha,\mathrm{MS}}$). Similarly, mean and RMS of the difference in 21-cm signal magnitude are $\langle\Delta |T_{21}|\rangle=\SI{48}{\milli\kelvin}$ and $\sqrt{\langle(\Delta |T_{21}|)^2\rangle}=\SI{60}{\milli\kelvin}$, respectively. The box-averaged 21-cm signal is $\SI{-207.3}{\milli\kelvin}$ without multiple scatterings and $\SI{-159.2}{\milli\kelvin}$ with multiple scatterings. Consequently, the relative difference (RMS) in terms of 21-cm signal between with and without multiple scatterings is $38.0\,\%$. We conclude that accounting for multiple scatterings can have a significant impact on the 21-cm signal. Multiple scatterings also have a substantial impact on the 21-cm power spectrum at large scales. We discuss this below in Section~\ref{ps}.

\subsection{Relative importance of various physical processes in setting the Ly\texorpdfstring{$\alpha$}{α} coupling}
\begin{figure*}
\centering
\includegraphics[width=0.65\linewidth]{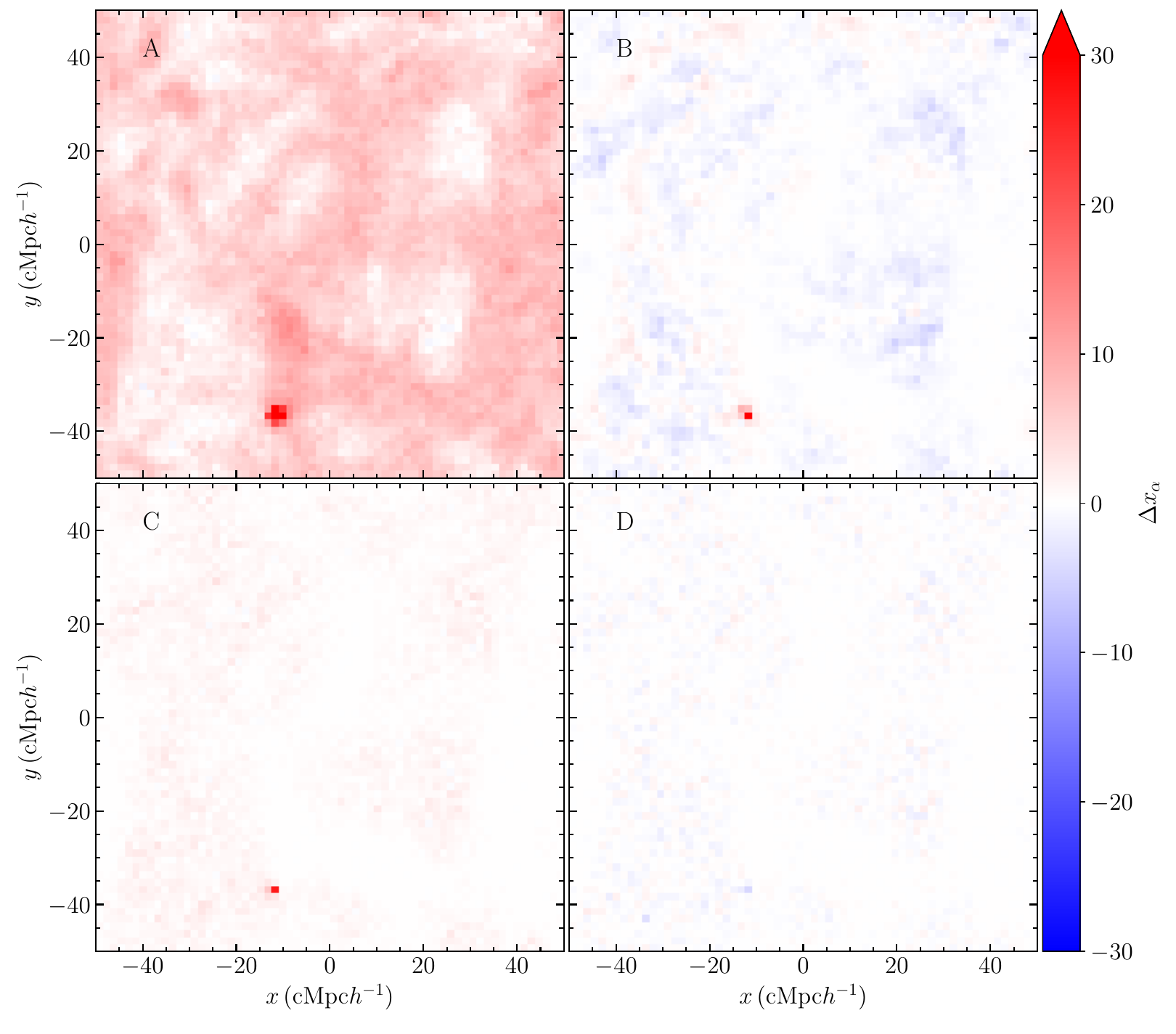}
\includegraphics[width=0.65\linewidth]{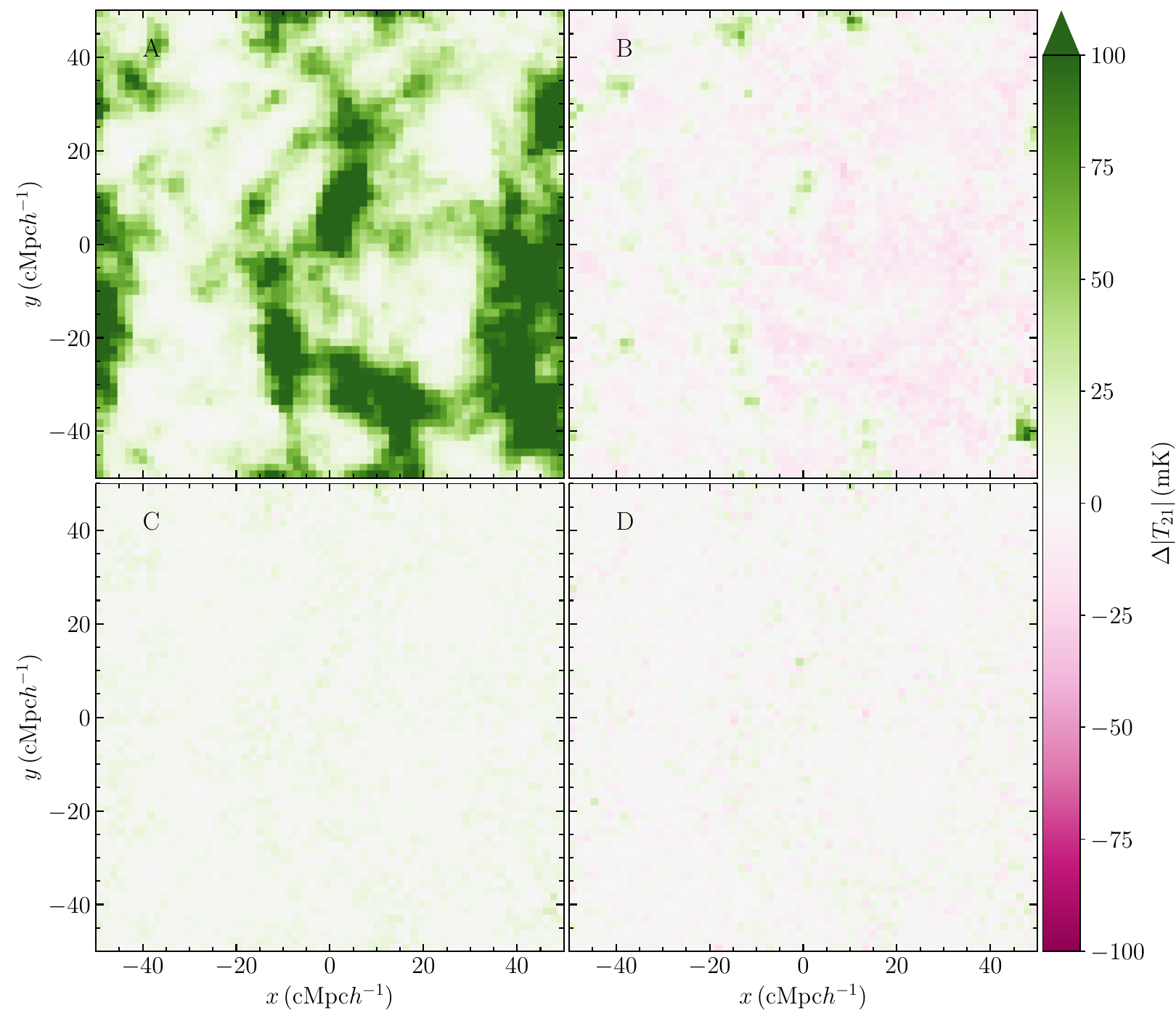}
\caption{The importance of different ingredients of Ly$\alpha$ radiative transfer for the computation of Ly$\alpha$ coupling $x_\alpha$ (top) and 21-cm signal $T_{21}$ (bottom). The colour shows the difference $\Delta x_{\alpha}$ in the Ly$\alpha$ coupling (top) and $\Delta |T_{21}|$ in the 21-cm signal (bottom) between each of our specially designed simulations A to D and fiducial simulation. The quantities $\Delta x_{\alpha}$ and $\Delta |T_{21}|$ are slices through the middle of the box perpendicular to the $\protect\cz$ axis of the simulation box. The difference $\Delta x_{\alpha}$ is given by $x_{\alpha,i}-x_{\alpha,0}$, where $x_{\alpha,0}$ is the Ly$\alpha$ coupling from our fiducial simulation, and $x_{\alpha,i}$ is the Ly$\alpha$ coupling in each of the simulations A to D. (A similar definition is used for $\Delta |T_{21}|$). The simulations A to D are described in Table~\ref{RTeffects}.}\label{compare_t21}
\end{figure*}

\citet[][LR99]{Loeb_1999} considered RT of Ly$\alpha$ photons emitted by a point source in a uniform, homogeneous and neutral medium undergoing a Hubble expansion but at a zero temperature. (When we write zero temperature we mean that the line profile is a Lorentzian in the gas frame because the thermal velocity is 0.) In this framework, because of zero thermal velocity of atoms and absence of recoil, scattering does not change the photon frequency. Therefore, frequency changes only as a result of redshifting.
 
To our knowledge, \citetalias{Zheng} were the first to consider the effect of multiple scatterings of Ly$\alpha$ photons in the context of the cosmological 21-cm signal. However, similar to \citetalias{Loeb_1999} the RT algorithm was applied to a homogeneous medium having no thermal motions. Moreover, only wing scatterings (applicable to photons having large frequency offset) were considered.

The seminal work by \citetalias{Semelin} introduced a 3D RT code which could be applied to real cosmological settings without recourse to assumptions like homogeneous density and zero temperature. They studied 3D RT for some idealised configurations such as a homogeneous medium with a point source at the centre, a high-density central clump at the centre or an isothermal density profile. \citet{Baek2009} applied the technology from \citetalias{Semelin} for the first time to real cosmological simulations. The period of cosmic history of interest in their work was the epoch of reionisation.

\citetalias{Semelin} and followup work by \citet{Baek2009} assumed isotropic scattering whereas in our work we consider the more accurate phase function \citep{Dijkstra08}. These authors have used acceleration schemes like core-skipping algorithms that we avoid in this work. When a photon is in its line core it undergoes a large number of scatterings over short distances. Often in MCRT codes, to speed up the computation, the photon is moved directly to the wings \citep{Ahn_2002}. In particular, \citet{Baek2009} trigger core-skipping when the photon is within $10\Delta\nu_\mathrm{D}$ of the $\nu_\alpha$ and add $\sim10^6$ scatterings `by hand' at the same location. Such an approach would miss the gas inhomogeneities, among these gas velocities being the dominant one. They account for the density and temperature inhomogeneities corresponding to the scatterings in the line core through the use of spectral distortions but still miss the gas velocity effects. We do an RT self-consistently throughout the line profile (a Voigt line) without core-skipping.

Another notable work that deals with Ly$\alpha$ RT is by \citet{Naoz}. They calculate the specific intensity due to wing scatterings by inserting a `correction' factor $f_{\mathrm{ws}}(n+1\to n,r_{z'z})$. This factor would serve as a correction for photons of frequencies between Ly$n$ and Ly$(n+1)$ when the photon was emitted at a redshift $z'$ and shifts into Ly$\alpha$ at $z$ such that the straight line distance is $r$ from the point of emission and the point where the photon becomes a Ly$\alpha$ photon.

\citet{Reis_2021, Reis_2022} further extended the approach taken by \citet{Naoz}. They fit a general form of a function to the histogram which describes the number of photons in the logarithmic bin at $r$. Similar to \citetalias{Loeb_1999} or \citetalias{Zheng} their MCRT procedure assumes a Lorentzian line profile. A Lorentzian line profile is an accurate description of the RT but only in the wings. The approach taken by \citet{Reis_2021, Reis_2022} does not account for the bulk and thermal motion of the gas, spatial fluctuations in the hydrogen density, anisotropy and recoil in the scattering process. Though these effects are subdominant in the wings, they become critically important in the line core.

Recently, \citet{Semelin2023} improve upon the previous work \citep{Semelin, Baek2009} by accounting for line core RT effects via the spectral distortion scheme by \citet{Hirata} along with an additional correction to include gas bulk velocity effects.

To re-emphasise, compared to previous works we account for all the RT effects self-consistently in the radiative transfer simulations itself throughout the line, i.e., core as well as wings.

In Fig.~\ref{new_results} we compared our results to a system which does not account for multiple scatterings at all. We now compare our results accounting for multiple scatterings of Ly$\alpha$ photons with results from specially designed simulations in which each time we invoke a specific RT assumption in effect, namely bulk motion, cosmological \ion{H}{i} density, recoil and scattering direction. Table~\ref{RTeffects} shows the different configurations against which we make our comparisons. The simulation details remain the same as above. Throughout the following discussion, we refer to the results of the right-hand-side panels of Fig.~\ref{new_results} as our fiducial results. 

The resulting Ly$\alpha$ coupling difference, $\Delta x_{\alpha}$, and difference in 21-cm signal magnitude, $\Delta |T_{21}|$, slices perpendicular to the $\cz$ direction are shown in Fig.~\ref{compare_t21}. We show $\Delta x_{\alpha}$ and $\Delta |T_{21}|$ in 4 different cases labelled A to D. We discuss each case in detail below, but Table~\ref{RTeffects} give the essential details. In Fig.~\ref{compare_t21}, $\Delta x_{\alpha}$ is defined as $x_{\alpha,i}-x_{\alpha,0}$, where $x_{\alpha,0}$ is our Ly$\alpha$ coupling predicted in fiducial model (shown in the right column of Fig.~\ref{new_results}) and $x_{\alpha,i}$ is that obtained for one of the A to D configurations. We define the quantity $\Delta |T_{21}|$ similarly.

In configurations which require a non-cosmological density, we replace the $n_{\ion{H}{i}}$ in all the cells of the simulation box to a constant value $\SI{545}{\metre^{-3}}$, which is the average neutral hydrogen density in our fiducial box at $z=13.2$.

Note that all of these assumptions refer only to the RT. So the assumption of uniform gas density employed in these numerical experiments will affect the RT computation of $x_\alpha(\bm{r},z)$, but when we compute the 21-cm signal, $T_{21}(\mathbfit{r},z)$, we still use the cosmological values $\delta(\bm{r},z)$ obtained from our \texttt{RAMSES} simulation.

We now discuss each configuration in detail.
\subsubsection*{A}
Configuration A differs from the fiducial model only in accounting of bulk motion, i.e., we set $\bm{v}_{\mathrm{b}}$ to 0. Even though the comoving frequency smoothly decreases as a result of redshifting (by ensuring small enough $\delta l$) the cell frame frequency can have large jumps from point to point in space due to varying bulk velocities. This may occasionally push the photon out of the line core thereby decreasing the number of scatterings. If $\bm{v}_{\mathrm{b}}=0$ the frequency decrease is smoother and the photons might stay in the line core for longer time resulting in higher number scatterings. This overpredicts the scatterings and hence Ly$\alpha$ coupling. The box-averaged $T_{21}$ is $\SI{-194.1}{\milli\kelvin}$ compared to our $\SI{-159.2}{\milli\kelvin}$. The RMS difference and the RMS relative difference in the 21-cm signal magnitude compared to fiducial model are $\SI{49}{\milli\kelvin}$ and $30.9\,\%$, respectively.

\subsubsection*{B}
In configuration B we use a uniform number density of hydrogen but everything else is same as in fiducial model. At first, the expression of rate of scatterings per atom, $P_\alpha$, suggests that it should be independent of number density (see equation~\ref{palpha}). However, the number density variations affect the photon trajectories and hence $P_\alpha$. Nevertheless, this is a sub-dominant effect as evident from Fig.~\ref{compare_t21} panel B. The box-averaged $T_{21}$ comes out to be $\SI{-156.9}{\milli\kelvin}$, which is close compared to our $\langle T_{21}\rangle=\SI{-159.2}{\milli\kelvin}$ and the RMS difference is $\SI{8.7}{\milli\kelvin}$. The RMS relative difference in the 21-cm signal magnitude compared to fiducial model is $5.5\,\%$.

\subsubsection*{C}
In this configuration everything is same as in the fiducial model except for recoil. Not including recoil somewhat overestimates $x_\alpha$. This is mostly because with recoil included, the photons lose energy faster compared to the case when recoil is excluded. As a result `non-recoiling' photons redshift out of the line core at a slower rate compared to recoiling photons. This allows non-recoiling photons to spend more time in the core thereby increasing the number of scatterings. Because we get a higher $x_\alpha$ we get a stronger 21-cm signal as evident in Fig.~\ref{compare_t21} panel C. The box-averaged $T_{21}$ comes out to be $\SI{-163.7}{\milli\kelvin}$ compared to our $\SI{-159.2}{\milli\kelvin}$. The RMS difference and the relative difference in the 21-cm signal magnitude compared to fiducial model are $\SI{5.5}{\milli\kelvin}$ and $3.5\,\%$, respectively.

Note that recoil is the physical mechanism which establishes a thermal equilibrium between the Ly$\alpha$ photons and the gas. It is this process which sets the colour temperature, $T_\alpha$, to the gas temperature \citep{Seon_2020}. When we calculate the spin temperature in configuration C we assume $T_\alpha=T_{\mathrm{k}}$.

\subsubsection*{D}
In this configuration everything is same as in fiducial model except for the scattering phase function. Here we use a uniform function as appropriate for isotropic scattering. The box-averaged $T_{21}$ in this configuration is $\SI{-159.7}{\milli\kelvin}$ compared to our $\SI{-159.2}{\milli\kelvin}$. The RMS difference and the RMS relative difference in the 21-cm signal magnitude compared to fiducial model are $\SI{4.4}{\milli\kelvin}$ and $2.8\,\%$, respectively.\\
[1\baselineskip]

The impact of bulk motion is the most significant compared to others for RT of Ly$\alpha$ photons and in establishing the 21-cm signal as suggested by the RMS difference values. This is also in agreement with the recent work by \citet{Semelin2023}. Thus, at the first level of approximation a cosmological $n_{\ion{H}{i}}$, recoil and anisotropy may be ignored when a box-averaged 21-cm signal is desired as the mean difference is small between fiducial model and B, C or D. The mean percentage difference in the 21-cm signal is not more than $3\,$\% in any of these configurations.
\begin{figure}
\centering
\includegraphics[width=1\linewidth]{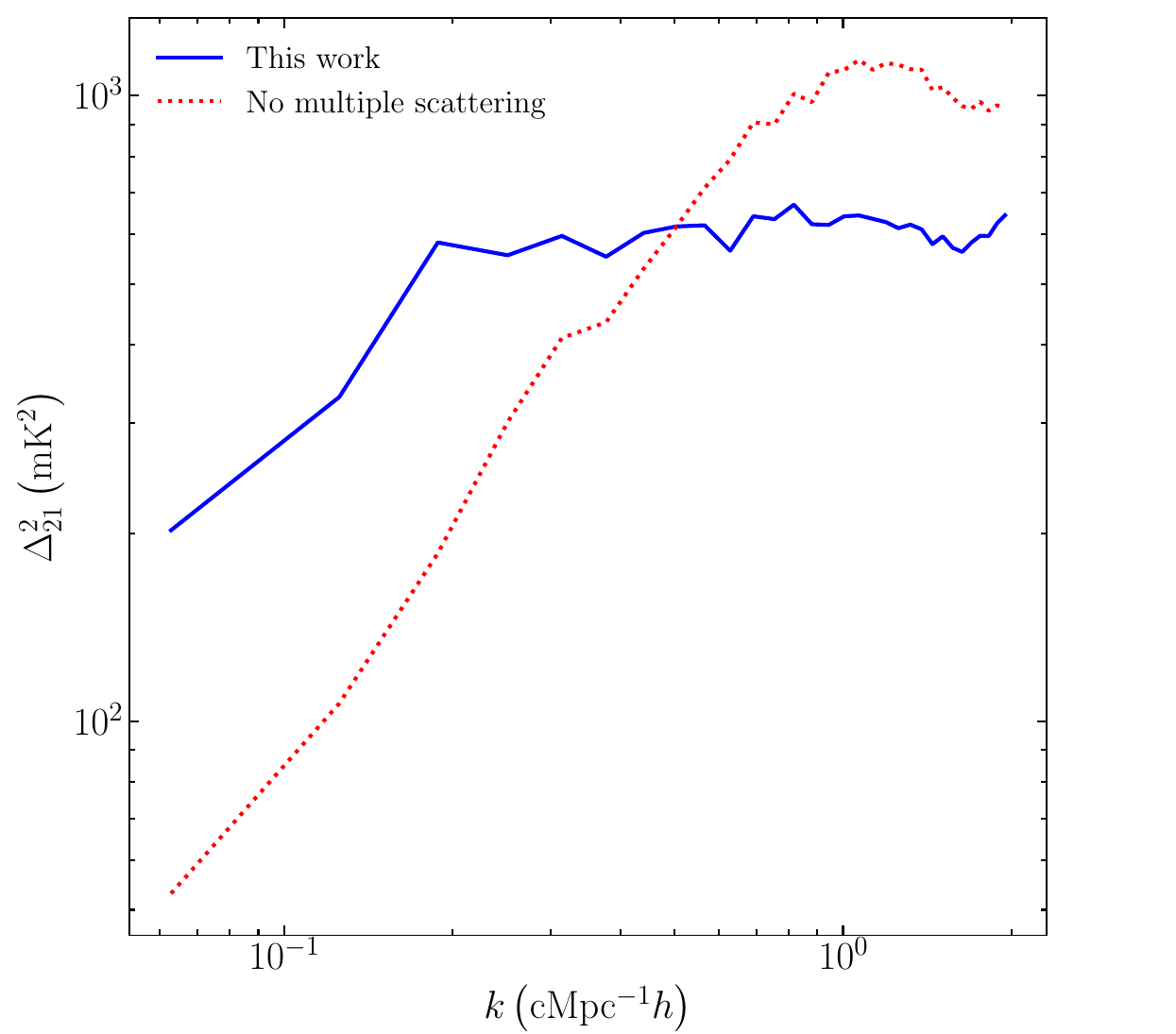}
\caption{The 21-cm power spectrum $\Delta^2_{21}$ for cosmic dawn like conditions ($z\sim13$). The blue-solid and red-dotted curves correspond to results accounting for multiple scatterings and no multiple scatterings, respectively; Fig.~\ref{new_results} shows the 21-cm maps for these.}\label{fig:ps}
\end{figure}

\subsection{21-cm power spectrum}\label{ps}
We next investigate how the fluctuations in 21-cm signal differ in cases with and without multiple scatterings. In our work there are three major contributors to the total 21-cm fluctuations, viz., gas temperature, density and Ly$\alpha$ scattering as evident from the 21-cm expression in equation~\eqref{DeltaT}. (As hydrogen is mostly neutral and uniform throughout, $x_{\ion{H}{i}}$ fluctuations have a negligible contribution. Similarly, the velocity gradient term has a negligible contribution to the total fluctuations.) However, as already mentioned, in the case with no multiple scatterings the Ly$\alpha$ coupling is saturated and as a result only gas temperature and density contribute to the fluctuations. See Fig.~\ref{fig:ps} where we show results in terms of $\Delta^2_{21}\,(\mathrm{mK}^2)$ as a function of comoving wavenumber $k\,(\mathrm{cMpc}^{-1}h)$.

\begin{figure}
\centering
\includegraphics[width=1\linewidth]{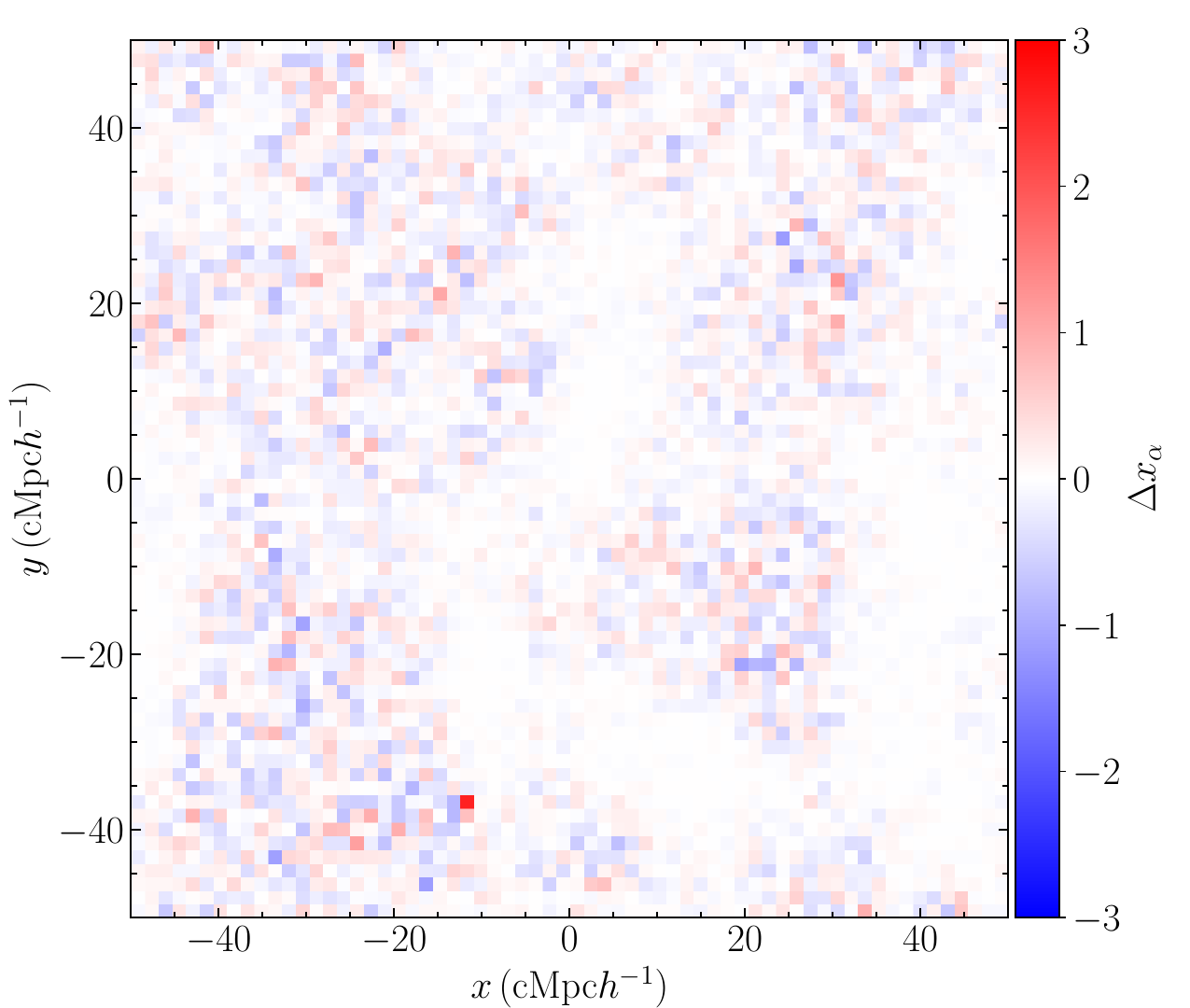}
\caption{Middle slice of $\Delta x_\alpha$ perpendicular to the $\protect\cz$ axis. We compare our fiducial run with $N_{\mathrm{mc}}=10^8$ against that done with $N_{\mathrm{mc}}=2\times10^8$. The mean difference $\Delta x_\alpha$ is $\sim\mathrm{10^{-4}}$ and RMS of difference is $\approx0.25$. The box-averaged 21-cm signal for higher MC photons comes out to be $\SI{-159.3}{\milli\kelvin}$ which is in an excellent agreement with $\SI{-159.2}{\milli\kelvin}$ for a lower MC photon count. This implies a convergence of our MCRT simulations in terms of number of MC photons. See text for more details.}\label{1e8vs2e8}
\end{figure}

The blue-solid curve corresponds to our fiducial simulations, the case of with multiple scatterings. The red-dotted curve corresponds to the case of no multiple scatterings. For case of with multiple scatterings, at $k=0.07\,\mathrm{cMpc}^{-1}h$ we have $\Delta^2_{21}\approx\SI{216.8}{\milli\kelvin^2}$, which is expected to be within the 68\,\% limits of the 21-cm power spectrum inferred by \textit{HERA} phase I observations at $z\sim13$ \citep{Abdurashidova_2022, Abdurashidova_2023}. For the case of no multiple scatterings, $\Delta^2_{21}(k)$ increases monotonically towards large $k$. 

\begin{figure*}
\centering
\includegraphics[width=0.9\linewidth]{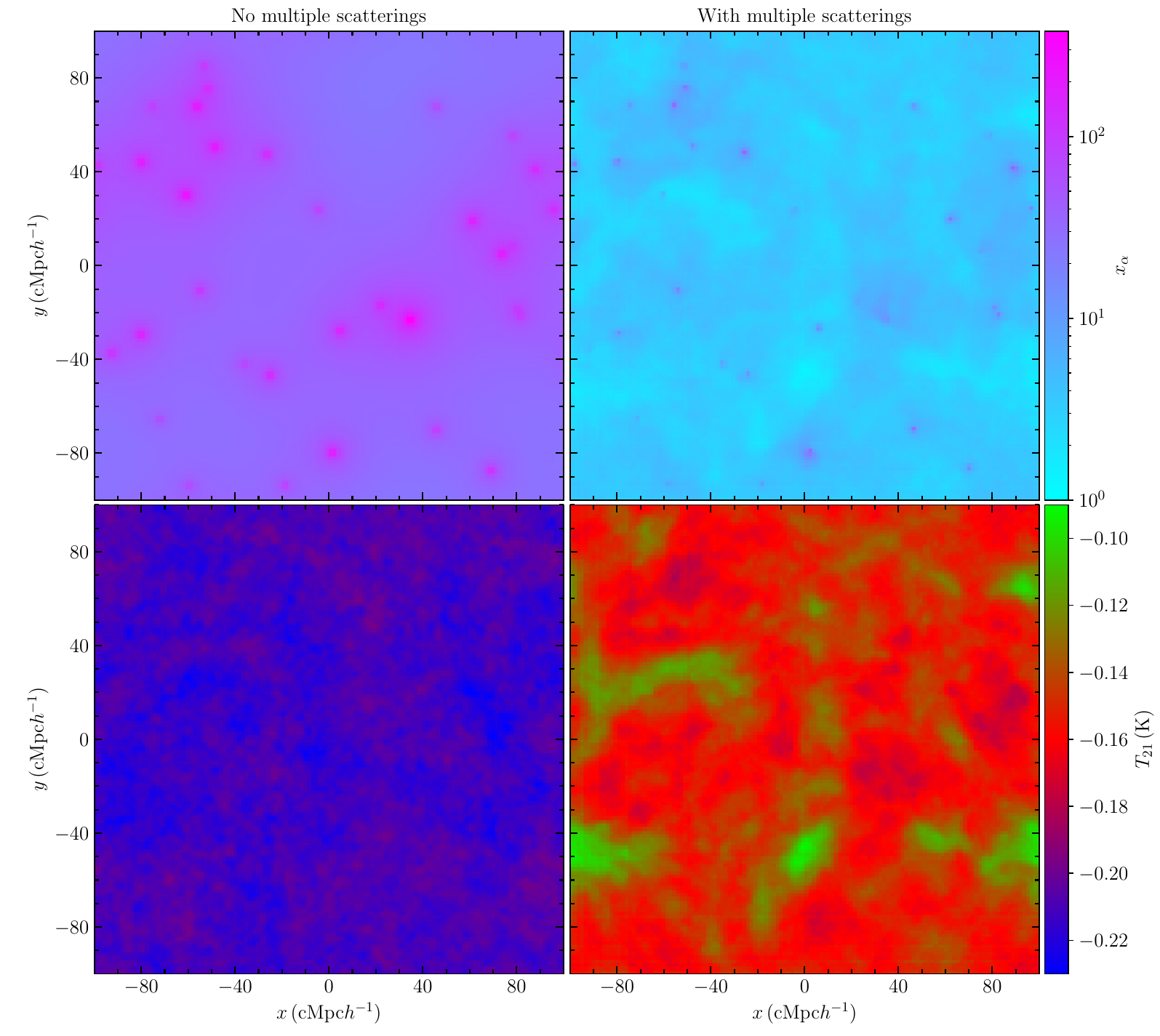}
\caption{Effect of the box size on our results. This figure has the same quantities as for Fig.~\ref{new_results}, but for a box of size $200\,\mathrm{cMpc}h^{-1}$ with $128^3$ cells. Note that the colour bar ranges are different than those in Fig.~\ref{new_results}. The box-averaged 21-cm signal for left- and right-hand sides are $\langle T_{21}\rangle=\SI{-211.7}{\milli\kelvin}$ and $\SI{-149.3}{\milli\kelvin}$, respectively. The redshift of simulation is $z\sim13$.}\label{200cMpc}
\end{figure*}

\section{Possible future improvements}\label{caveats}
The above results indicate the effect that RT has on Ly$\alpha$ coupling at cosmic dawn. The results also help us understand the relative importance of various gas properties in setting the Ly$\alpha$ coupling. Nonetheless, we can now identify possible future improvements. These include improvements in the simulation box size, number of MC photons, spatial and mass resolution, and the MCRT stopping criterion. We now discuss each of these points.

Perhaps the most important factor which decides convergence of RT simulations is the number of MC photons $N_{\mathrm{mc}}$. As is generally true for MC process, an MCRT simulation result would suffer from sampling noise. Following \citet{Semelin2023}, the relative error in our Ly$\alpha$ coupling, when a broad Voigt-like line profile is adopted, can be estimated to be $\eta=\sqrt{N_\mathrm{cell}/N_\mathrm{mc}}$, where $N_\mathrm{cell}$ is the number of cells in the simulation box. We did our main runs with $N_\mathrm{cell}=64^3$ and $N_\mathrm{mc}=10^8$ which gives $\eta\approx5\,$\%. This is a sufficiently small MC noise. A more realistic simulation would have a higher hydrodynamic resolution, such as $2048^3$ cells. However, in order to get a similar or lower level of MC noise one would require $N_\mathrm{mc}\gtrsim 3\times10^{12}$. A simulation with these many number of cells and photons can be prohibitively costly and thus, has been left for future. Nevertheless, we run a simulation with $2\times10^8$ MC photons and compare the results with our fiducial model for which we used $10^8$ MC photons. 

Figure~\ref{1e8vs2e8} shows the difference, $\Delta x_\alpha$, in Ly$\alpha$ coupling using $10^8$ and $2\times10^8$ MC photons. We define $\Delta x_\alpha$ as $x_{\alpha,1}-x_{\alpha,0}$, where $x_{\alpha,0}$ and $x_{\alpha,1}$ are the Ly$\alpha$ couplings with $10^8$ (the fiducial run, shown in the right-hand side column of Fig.~\ref{new_results}) and $2\times10^8$ MC photons, respectively. The mean and RMS of the difference over the complete domain are $\langle\Delta x_\alpha\rangle=-4.9\times10^{-4}$ and $\sqrt{\langle(\Delta x_\alpha)^2\rangle}=0.25$, respectively. The box-averaged 21-cm signal comes out to be $\langle T_{21}\rangle=\SI{-159.3}{\milli\kelvin}$ which is in an excellent agreement compared to $\SI{-159.2}{\milli\kelvin}$ when $10^8$ MC photons are used. This is quite promising indicating that for fixed hydrodynamics, our radiative transfer is not affected by the number of MC photons, thus implying convergence.

We next investigate the effect of simulation box size. Consider a box of side length $200\,\mathrm{cMpc}h^{-1}$ with $128^3$ cells (smoothed down from $1024^3$). This set-up differs from our fiducial model in terms of box length and number of cells but has the same resolution of $97\,\mathrm{ckpc}h^{-1}$. Since the simulation box is different, we rerun our halo finder to find the sources and luminosity. We have a total of 33 haloes (cf. our fiducial configuration where we had 34 haloes) and luminosity, $L=\SI{2.8e59}{\second^{-1}}$. We show results in Fig.~\ref{200cMpc}. The box-averaged 21-cm signal without and with multiple scatterings is $\SI{-211.7}{\milli\kelvin}$ and $\SI{-149.3}{\milli\kelvin}$, respectively. We used a different seed for initial conditions, so that we have a different density and temperature distribution compared to that shown in Fig.~\ref{den}. As we have increased the number of cells but kept the MC photon count the same, this result suffers from a larger MC sampling noise. We note that qualitatively the results remain the same and we see here similar differences as for the case of a $100\,\text{cMpc}h^{-1}$ box shown in Fig.~\ref{new_results}. The Ly$\alpha$ coupling has a uniform and isotropic distribution around the sources in the case of no multiple scatterings but not in the case of multiple scatterings. As in the situation with $100\,\text{cMpc}h^{-1}$ box the Ly$\alpha$ coupling is higher in the case of no multiple scatterings than with multiple scatterings. This leads to a saturated spin temperature on the left with the 21-cm signal having a direct correspondence with the gas temperature.

Third, the spatial and mass resolution of the simulation can also be improved. Our fiducial simulation has a spatial resolution of $97\,\mathrm{ckpc}h^{-1}$ and the minimum halo mass resolved by the simulation is $2.89\times10^9\,\mathrm{M}_{\odot}h^{-1}$. It remains to be understood how unresolved gas density structures affect the Ly$\alpha$ RT. The halo mass resolution also should ideally extend down to the uncertain star formation threshold at cosmic dawn. While the technology presented in this paper is now potentially capable of answering these questions, the computational expense involved forces us to leave resolution improvements to future work.

Finally, we used the stopping criterion of $x_{\mathrm{cell,crit}}=-8$. As mentioned in Section~\ref{sec:sc}, for this choice of critical frequency, the critical cross-section is $\sim10^{-4}$ times the central cross-section. By running test cases with successively higher $-x_{\mathrm{cell,crit}}$ (at fixed $N_{\mathrm{mc}}$, box size, and resolution) we found that going beyond $x_{\mathrm{cell,crit}}\approx-8$ to conditions such as $x_{\mathrm{cell,crit}}=-16$ has less than a per cent impact on the 21-cm signal.

By bringing greater realism in the treatment of gas physics, this work provides a starting point for a more accurate computation of Ly$\alpha$ coupling and the 21-cm signal, keeping the above considerations in mind.

\section{Conclusions}\label{Conc}
In this work we have set up the technology to study multiple scatterings of Ly$\alpha$ photons for the computation of tomographic cosmological 21-cm signal at cosmic dawn. Using the AMR code \texttt{RAMSES} we performed hydrodynamical simulations to obtain the cosmological boxes giving us the density, temperature and bulk velocity of the gas. We post-processed our simulation with the \texttt{RASCAS} code to compute the propagation of Ly$\alpha$ photons with multiple scatterings using a Monte Carlo radiative transfer simulation. We modified \texttt{RASCAS} to account for the cosmological redshifting of photons and to compute the scattering rate using a path-based method. In contrast with previous works, we also account for recoil and anisotropic scattering, and more importantly, do not use core-skipping algorithms.

We investigate the role of multiple scatterings of Ly$\alpha$ photons due a finite spread of line profile against the traditional computation which implicitly assumes a Dirac-delta line profile. We also study the role played by different gas properties such as cosmological number density and bulk velocity in establishing the Ly$\alpha$ coupling and the 21-cm signal at cosmic dawn. Our main findings in this work are as follows.
\begin{enumerate}
\item[1)] Ly$\alpha$ coupling and consequently the 21-cm signal differ significantly when multiple scatterings are taken into account compared to the traditional calculations where no multiple scatterings are considered. The 21-cm signal in our fiducial simulation differs from the results of a traditional calculation by about 38\,\% (RMS). Our treatment brings greater realism in how intergalactic gas physics is modelled in the computation of the Ly$\alpha$ RT.

\item[2)] We do Ly$\alpha$ RT self-consistently taking into account the medium inhomogeneities throughout the line profile, core as well as wings, for the first time. Previous work in the literature have either ignored the medium inhomogeneities or those that improve upon this adopt core-skipping algorithms with certain recipes to capture the line-core effects.

\item[3)] We investigate the relative importance of the gas bulk motion, cosmological \ion{H}{i} density distribution, anisotropic scattering, and recoil, for Ly$\alpha$ RT. We conclude that the gas bulk motion is the most important of these gas properties.

\item[4)] We find that Ly$\alpha$ RT increases the 21-cm power spectrum at small scales and decreases at large scales compared to the traditional computations.
\end{enumerate}

The global 21-cm signal is a promising probe for the thermal state of the universe at cosmic dawn. Accurate inferences from its measurements are valuable for a range of astrophysical and cosmological processes such galaxy formation, initial mass function of the first stars, nature of dark matter among others. This work offers an understanding of the various radiative transfer effects in increasing the accuracy in the modelling of Lyman-$\alpha$ coupling and hence the 21-cm signal. This work paves the way forward for better modelling and inferring from future experiments such as \textit{REACH}.

\section*{Acknowledgements}
We thank the referee Ben\^oit Semelin for constructive and educational comments that helped improve this paper. It is a pleasure to acknowledge discussions with members of the Radio Experiment for the Analysis of Cosmic Hydrogen (\textit{REACH}) collaboration. GK gratefully acknowledges support by the Max Planck Society via a partner group grant. GK is also partly supported by the Department of Atomic Energy (Government of India) research project with Project Identification Number RTI 4002. We specially thank Thomas Gessey-Jones, Oliver Hahn, Mladen Ivkovic, Harley Katz, Sergio Martin-Alvarez, Leo Michel-Dansac, Kwang-il Seon, Aaron Smith and Yuxuan Yuan for useful discussions. We also thank the \texttt{yt} and \texttt{RAMSES} community for helping with the codes.

\section*{Data Availability}
The modified version of \texttt{RASCAS} developed by us for this work, along with the parameter files to reproduce the main results of this work will be made public soon. Additional codes to analyse the outputs, and parameter files to run \texttt{monofonIC} and \texttt{RAMSES} are available at \url{https://github.com/shikharmittal04/lyman-a}. 



\bibliographystyle{mnras}
\bibliography{Biblo} 


\appendix
\section{Code Tests}\label{test_rascas}
In this appendix we present some tests of our version of the \texttt{RASCAS} code. The first two tests are the standard spectra reproducible by a non-Hubble-expanding version of the code, while the remaining three are the newer tests we did for testing our modified code for the scattering rate and/or Hubble expansion.

Unless stated otherwise in all of the following tests we do not account for the bulk motion of the scatterers or recoil, and assume an isotropic scattering. Because of the latter condition the phase function is simply given by $\mathcal{P}(\mu)=1$ as opposed to equation~\eqref{phase} which we have used in the main results of this paper. Finally, note that we do \emph{not} assume periodic boundary conditions for these tests and stop the photon propagation when it just crosses the domain boundary. We run all our tests on a $256^3$ grid box.

\subsection{Effect of recoil on the output spectrum from a static medium}\label{rec}
In this test we investigate how the recoil affects the output spectrum from a uniform, non-absorbing, static, optically thick sphere with a point source sitting at the centre emitting at the Lyman-$\alpha$ frequency. The analytical solution in the absence of recoil is given by \citep{Dijkstra_2006}
\begin{equation}
J(x)=\sqrt{\frac{\pi}{6}}\frac{1}{4a\tau_0}\frac{x^2}{1+\cosh\left(\sqrt{\frac{2\pi^3}{27}}\frac{|x|^3}{a\tau_0}\right)}\,,\label{sphere_spec}
\end{equation}
valid for a large line centre optical depth from the centre of sphere to its boundary, $\tau_0=N_{\ion{H}{i}}\sigma(\nu_\alpha)$. Here $x$ is the frequency deviation form the line centre in Doppler units and $a$ is the Voigt parameter defined as the ratio of natural to Doppler line width. For Ly$\alpha$ scattering off hydrogen atoms
\begin{equation}
a=4.71\times10^{-4}\left(\frac{T}{\SI{e4}{\kelvin}}\right)^{-1/2}\,,
\end{equation}
and
\begin{equation}
\tau_0=8.3\times10^6\left(\frac{N_{\ion{H}{i}}}{\SI{2e24}{\metre^{-2}}}\right)\left(\frac{T}{\SI{2e4}{\kelvin}}\right)^{-1/2}\,,\label{od}
\end{equation}
for a uniform temperature $T$. The column density, $N_{\ion{H}{i}}$, is set by the number density of hydrogen and the system size, in this case the sphere radius $(r)$, so that $N_{\ion{H}{i}}=n_{\ion{H}{i}}r$.

The spectrum given by equation~\eqref{sphere_spec} is normalised such that
\begin{equation}
\int_{-\infty}^{\infty}J(x)\,\ud x=\frac{1}{4\pi}\,.\label{normal}
\end{equation}

\begin{figure}
\centering
\includegraphics[width=1\linewidth]{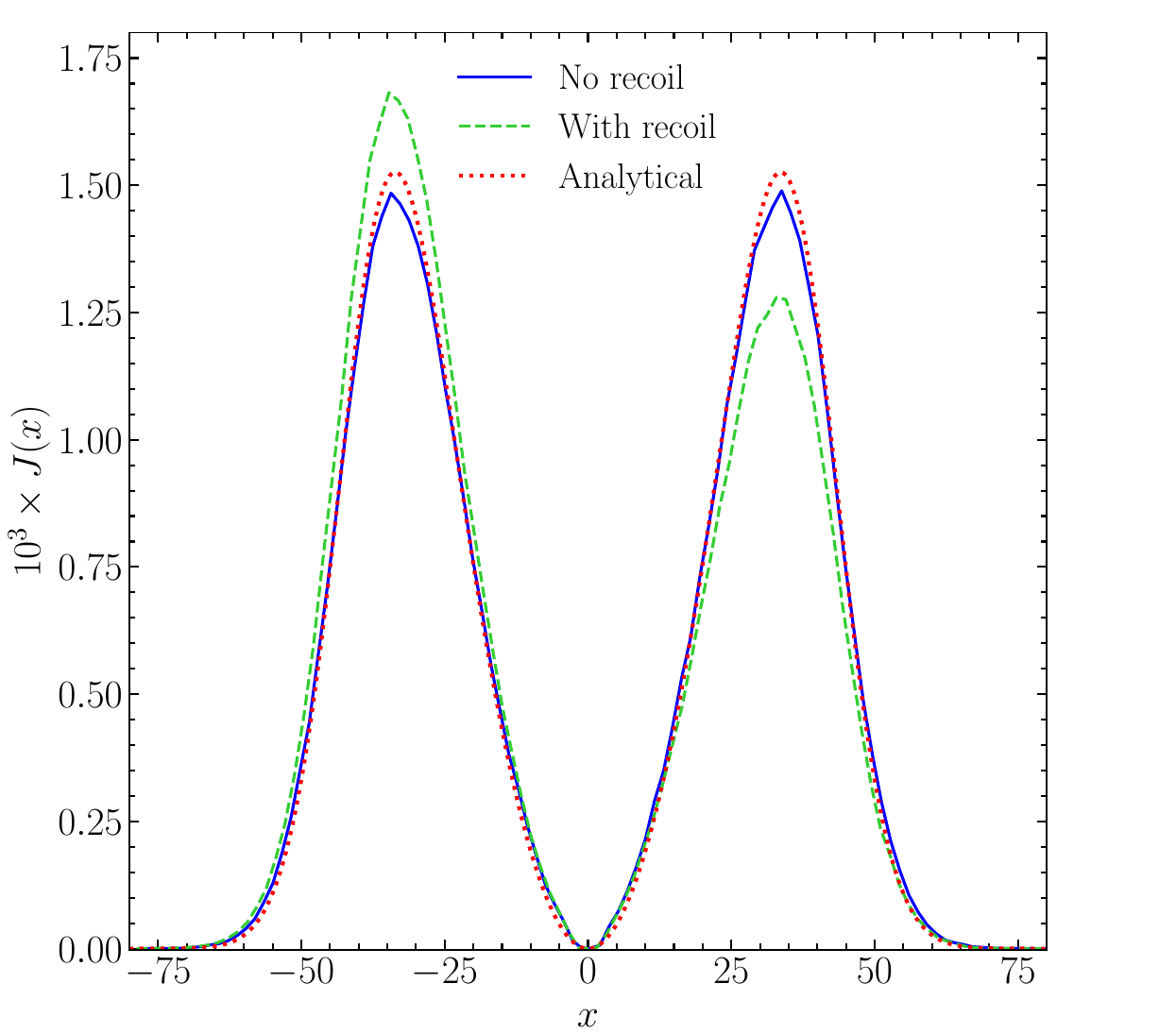}
\caption{The figure shows the output spectrum normalised to $1/4\pi$ from a homogeneous static sphere with a point source emitting from the centre at Ly$\alpha$ frequency. The medium temperature and the line centre optical depth are $\SI{100}{\kelvin}$ and $\tau_0=10^7$, respectively. The red dotted curve shows the analytical solution \citep{Dijkstra_2006} applicable to a medium with large optical depths such as in this case. Blue-solid curve shows the output predicted by running a \texttt{RASCAS} simulation when no recoil is taken into account. The agreement between the two curves shows an agreement between the analytical solution and numerical simulation. With recoil the spectrum is asymmetric as shown by the green-dashed curve.}\label{fig:rascstest1}
\end{figure}

Figure~\ref{fig:rascstest1} shows our results. We first note that the simulation result without recoil is in an excellent agreement with the analytical solution, as also established by \citet{rascas}. More importantly the figure illustrates the effect of recoil on the photons. In a static Universe and in the absence of recoil a photon is equally likely to get a positive and a negative Doppler shift on every scattering because of random thermal motion of scatterers. However, with recoil taken into account the frequency shifts are no longer unbiased; photons lose a small amount of energy to the scatterers. Hence, there are more photons with low energy which explains the enhanced left-hand-side peak.

For interested readers the following are our set-up parameters. The medium is at $T=\SI{100}{\kelvin}$, number density of neutral hydrogen is $n_{\ion{H}{i}}=\SI{3.477e17}{\metre^{-3}}$ and radius of the sphere is $r=\SI{4.9e5}{\metre}$ so that $N_{\ion{H}{i}}=\SI{1.7e23}{\metre^{-2}}$ and the line centre optical depth is $\tau_0=10^7$. We run our simulation for $10^6$ Monte Carlo (MC) photons. We normalise the resulting numerical histogram, using 100 bins, according to equation~\eqref{normal}.

\subsection{Radially- vs Hubble-expanding medium}\label{laursen}
\begin{figure}
\centering
\includegraphics[width=1\linewidth]{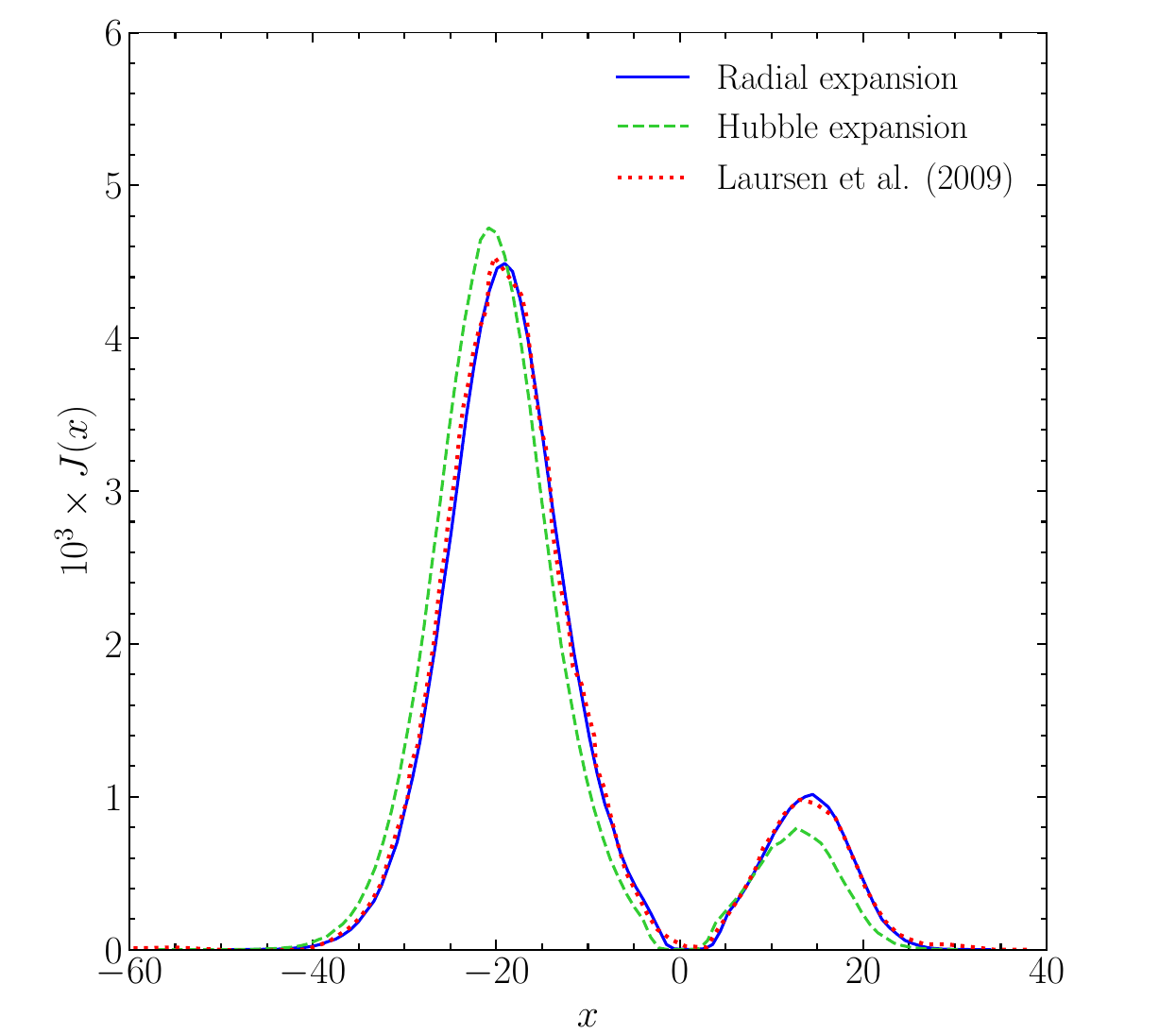}
\caption{We reproduce the result of Fig.~8 of \citet{Laursen_2009} and also illustrate how a true Hubble-expanding medium differs from a medium in which the atoms are given an additional Hubble-flow-like velocity field. Our set-up has a point source at the centre of our sphere emitting at Ly$\alpha$ frequency. The medium temperature, column density and the edge velocity are $\SI{e4}{\kelvin}$, $N_{\ion{H}{i}}=\SI{2e24}{\metre^{-2}}$ and $v_{\mathrm{max}}=\SI{20}{\kilo\metre\second^{-1}}$, respectively. The red-dotted curve shows the result from previous literature, the blue solid curve shows the result from a medium given a radial dilation and the green dashed curve represents output from a truly Hubble-expanding medium. Hubble expansion results in a higher frequency shift because of which the green curve is more leftwards than blue one.}\label{fig:rascstest2}
\end{figure}

Here we highlight the difference between output spectra from a medium simply given a radial dilation vs that from a Hubble-expanding medium. Our radially-expanding medium is similar to the previous configuration except this time the medium has an additional bulk motion. The velocity increases linearly with radius from 0 at the centre of sphere to $v_\mathrm{max}=\mathcal{H}r$ at the outermost shell, just as in the set-up of \citet{Laursen_2009}. In such a system the photons that are initially Doppler shifted to higher frequency due to random thermal motion of scatterers, on reaching larger radii (when radially outwards velocity becomes large) will appear closer to the line centre to scatterers. These photons have a high scattering rate compared to the initially redder photons which become even redder when reaching larger radii. Thus, more redder photons escape the boundary than the bluer ones which results in an asymmetric spectrum. We firstly note that the output from this system matches with that of \citet{Laursen_2009} as shown in Fig.~\ref{fig:rascstest2}.

Let us now consider Hubble-expanding medium. While equation~\eqref{redshift} at first seems to suggest redshifting can be captured by Doppler shifting the photon by giving the scatterer precisely the Hubble velocity, it is only applicable when the mean free path is small. An alternative and perhaps the best way to understand why the two cases are different is through the following thought experiment. Let us label the two cases as `radial' version and `Hubble' version for ease of discussion. Consider a very rare scenario where a photon moves from the centre to boundary without scattering even once. In radial version photon will exit the sphere with the same frequency as its emission frequency because it never touched an atom which could have caused a frequency shift. But in Hubble version the photon undergoes a cosmological redshift which has nothing to do with the presence of atoms and hence the photon exits with a lower frequency than it originally started with \citep{Dijkstra_2006}.

From the above discussion it is clear that radially-expanding and a Hubble-expanding medium will produce different outputs. As evident from Fig.~\ref{fig:rascstest2} the output spectrum from a Hubble-expanding medium (green dashed curve, labelled `Hubble expansion') is slightly more shifted compared to that from a radially expanding medium (blue solid curve, labelled `Radial expansion') in accordance with our expectation. Additionally, we conclude that to capture the cosmological redshifting, it is better to work in a comoving frame where one redefines photon frequency in gas frame (according to equation~\ref{redshift}) over predetermined small distances traversed by the photon while in free propagation. 

For interested readers the following are our set-up parameters. The medium has a uniform neutral hydrogen density of $n_{\ion{H}{i}}=\SI{3.84e3}{\metre^{-3}}$, a constant temperature of $T=\SI{e4}{\kelvin}$ and a radius of $\SI{1.69e-2}{\mega\parsec}$. The column density and line centre optical depth are $N_{\ion{H}{i}}=\SI{2e24}{\metre^{-2}}$ and $\tau_0\approx1.2\times10^7$, respectively. For $\mathcal{H}=\SI{1184.23}{\kilo\metre\second^{-1}\mega\parsec^{-1}}$ the maximum velocity -- which is at the edge of the sphere -- is $\SI{20}{\kilo\metre\second^{-1}}$. We run our simulation for $10^6$ MC photons. We normalise the resulting numerical histogram, using 100 bins, according to equation~\eqref{normal}. 

\subsection{Scattering rate in a non-Hubble-expanding medium}
Here we run our first test for the scattering rate per atom by comparing with the results from \citet{Seon_2020}. \texttt{RASCAS} version introduced by \citet{rascas} did not have the $P_{\alpha}$ calculation capabilities. We have introduced this in our version.

The set-up is similar to the previous cases where we have a monochromatic source emitting at Ly$\alpha$ frequency from the centre of sphere. The column density is $\SI{1.69e19}{\metre^{-2}}$ and the temperature is $\SI{100}{\kelvin}$ giving us a line centre optical depth of $\tau_0=10^3$. We run our simulations for two sub-cases -- i) where we have a static medium, and ii) where medium is expanding outwards (not Hubble) with the outermost velocity being $\SI{10}{\kilo\metre\second^{-1}}$. We make a radial profile of $P_{\alpha}$ to facilitate a comparison with previous work. For our numerical set-up we used $10^7$ MC photons and 80 histogram bins for the radial profile.

\begin{figure}
\centering
\includegraphics[width=1\linewidth]{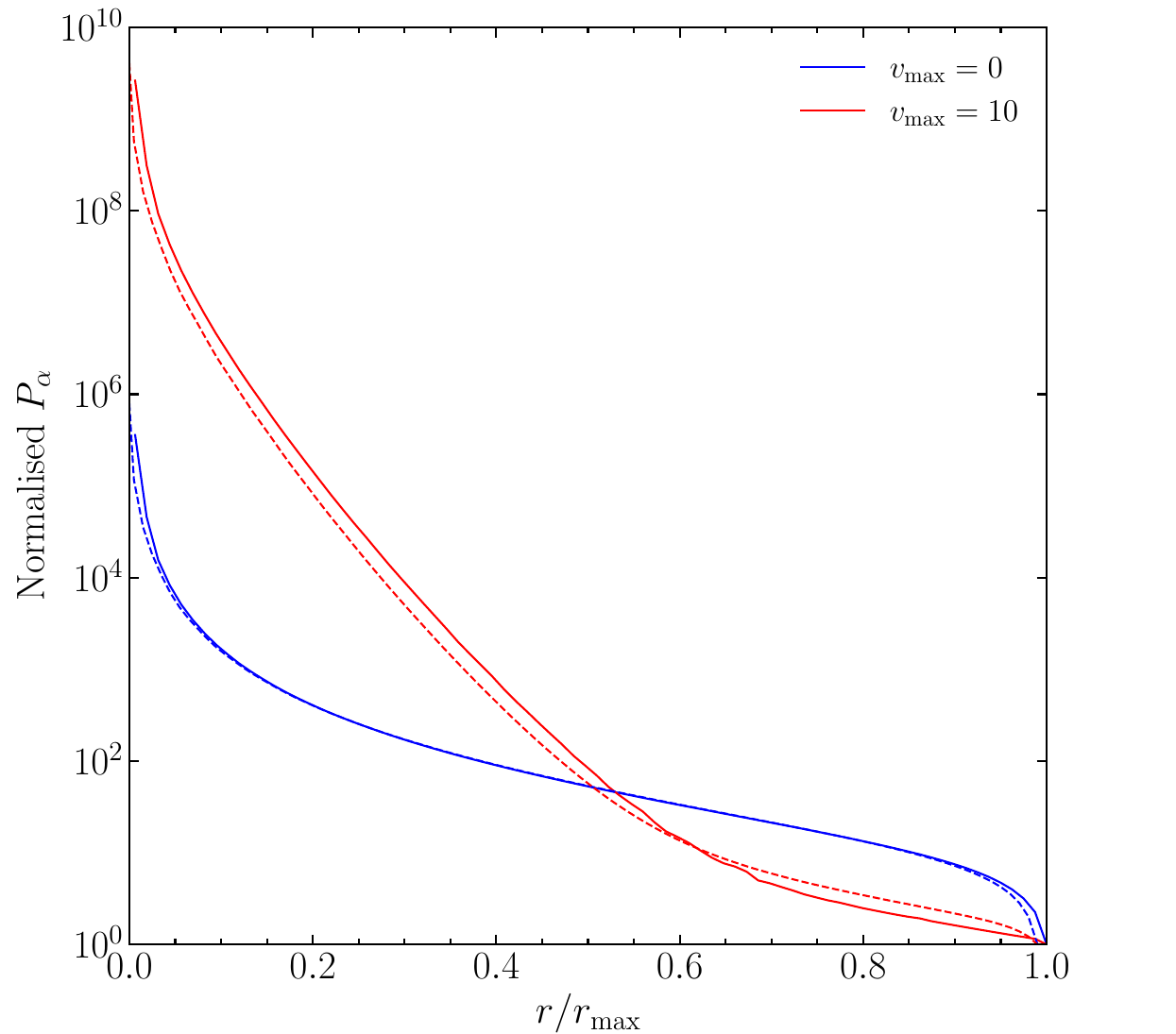}
\caption{Spherically-averaged radial profile of scattering rate $P_{\alpha}$ in a medium of uniform density and temperature ($\SI{100}{\kelvin}$) with a single source emitting at Ly$\alpha$ frequency from the centre. The line centre optical depth is $\tau_0=10^3$. The blue and red curves correspond to a static and a radially expanding media, respectively. $v_{\mathrm{max}}$ indicates the maximum velocity observed at the spherical boundary in units of $\si{\kilo\metre\second^{-1}}$. The solid curves, results from our own simulations, match reasonably well with the dashed curves which are the results by \citet{Seon_2020}.}\label{fig:sk20}
\end{figure}

As evident from Fig.~\ref{fig:sk20} we match reasonably well with results from previous literature. As the source strength is irrelevant for this test case we show a normalised version of $P_{\alpha}$ such that $P_{\alpha}=1$ at the farthest distance from the source. Also, we normalise the distance with the sphere radius. Solid and dashed curves are ours and their results, respectively.

\subsection{Spectrum in a Hubble-expanding medium}\label{lr99}
Here we reproduce the analytical solution given by \citet[][LR99]{Loeb_1999}. In this set-up the spherical medium undergoes a Hubble expansion and is at absolute zero. The consequence of the latter is that the gas atoms do not have any thermal motion and the line profile of the Ly$\alpha$ photons is simply given by a Lorentzian as follows
\begin{equation}
\sigma_{\mathrm{L}}(\nu)=\frac{3\lambda_\alpha^2}{2\pi}\frac{\tilde{A}^2}{(\nu-\nu_\alpha)^2+\tilde{A}^2}\,,
\end{equation}
where $\tilde{A}=A_{21}/(4\pi)$. However, to be exactly consistent with \citetalias{Loeb_1999} radiative transfer physics we use the wing approximation to the above form, i.e.,
\begin{equation}
\sigma_{\mathrm{L}}(\nu)\approx\frac{3\lambda_\alpha^2}{2\pi}\frac{\tilde{A}^2}{(\nu-\nu_\alpha)^2}\,.
\end{equation}
We place our source at the centre of the sphere emitting at the frequency of Ly$\alpha$, i.e., $\nu=\nu_\alpha$. Because of the absence of thermal motion and recoil, frequency is not affected by scattering but changes merely due to redshifting. Note that contrary to the previous set-up, this time we have a true Hubble expansion.

The analytical solution for the set-up described above is given by
\begin{equation}
\tilde{J}(\tilde{\nu},\tilde{r})=\frac{1}{4\pi}\left(\frac{9}{4\pi\tilde{\nu}^3}\right)^{3/2}\exp\left(-\frac{9\tilde{r}^2}{4\tilde{\nu}^3}\right)\,,\label{analytic}
\end{equation}
where $\tilde{r}=r/r_{\star}$ and $\tilde{\nu}=(\nu_\alpha-\nu)/\nu_{\star}$, where $\nu_{\star}$ is a convenient frequency scale introduced by \citetalias{Loeb_1999} for which a photon initially at frequency $\nu_\alpha-\nu_\star$ has an optical depth 1 out to infinity. Similarly, $r_\star$ is the proper distance from the source where the Hubble expansion produces a frequency shift of $\nu_\star$.

For our numerical set-up we prepare a 2D histogram of 1000 bins in both, $\tilde{r}$ and $\tilde{\nu}$ space. We increment the bin value by 1 corresponding to photon's current position and frequency as the photon is made to propagate over small distances. We run our simulations at $z=10$ for which $\nu_\star=\SI{1.2e13}{\hertz}$, $r_\star=\SI{3.22e22}{\metre}$ and assigned uniform density of $\SI{254}{\metre^{-3}}$. We normalise our numerical histogram such that the peak of our numerical result coincides with that of analytical curve. We use $10^7$ MC photons for this test.

\begin{figure*}
\centering
\begin{subfigure}{0.48\textwidth}
\includegraphics[width=1\linewidth]{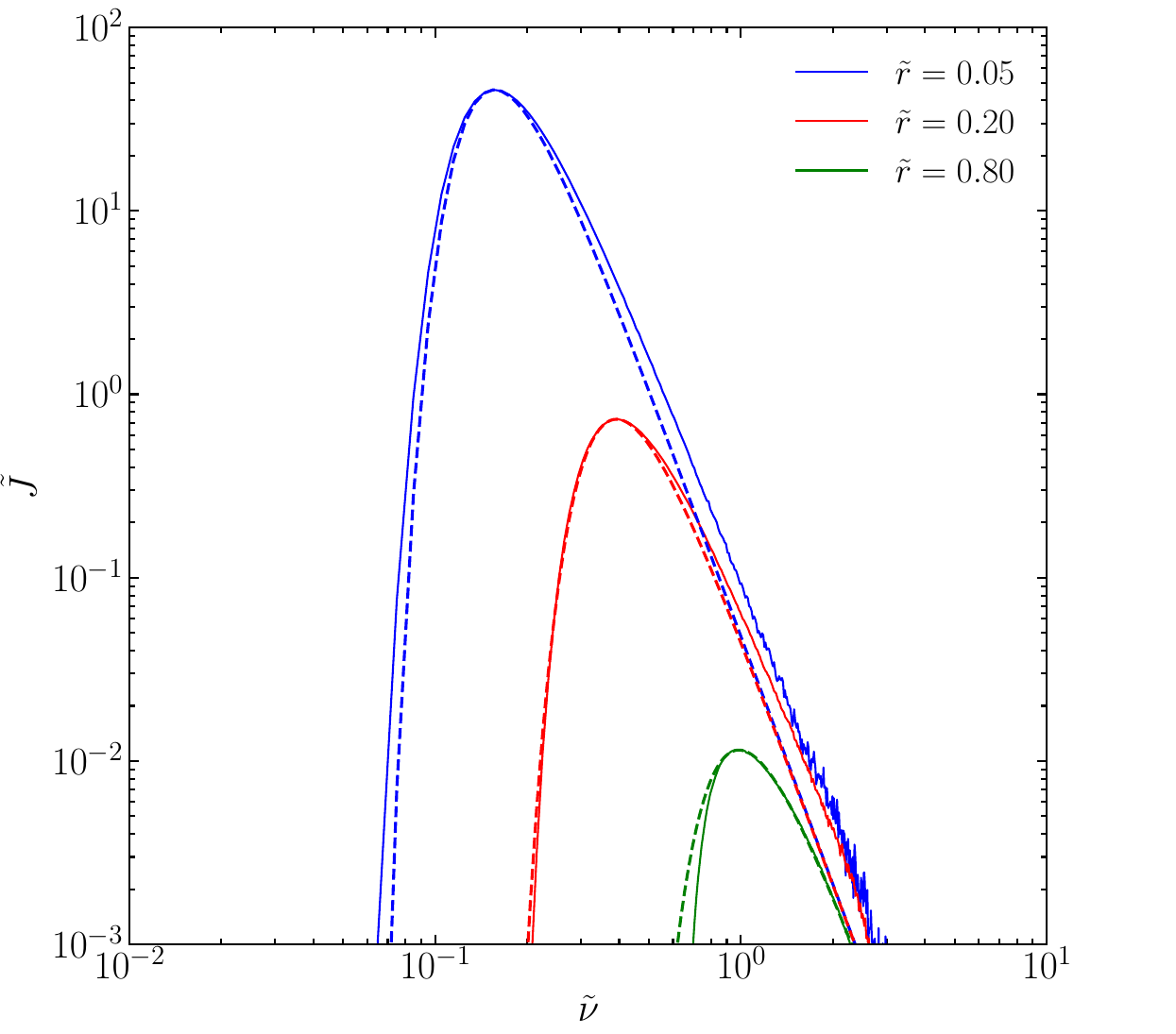}
\end{subfigure}%
\begin{subfigure}{0.48\textwidth}
\includegraphics[width=1\linewidth]{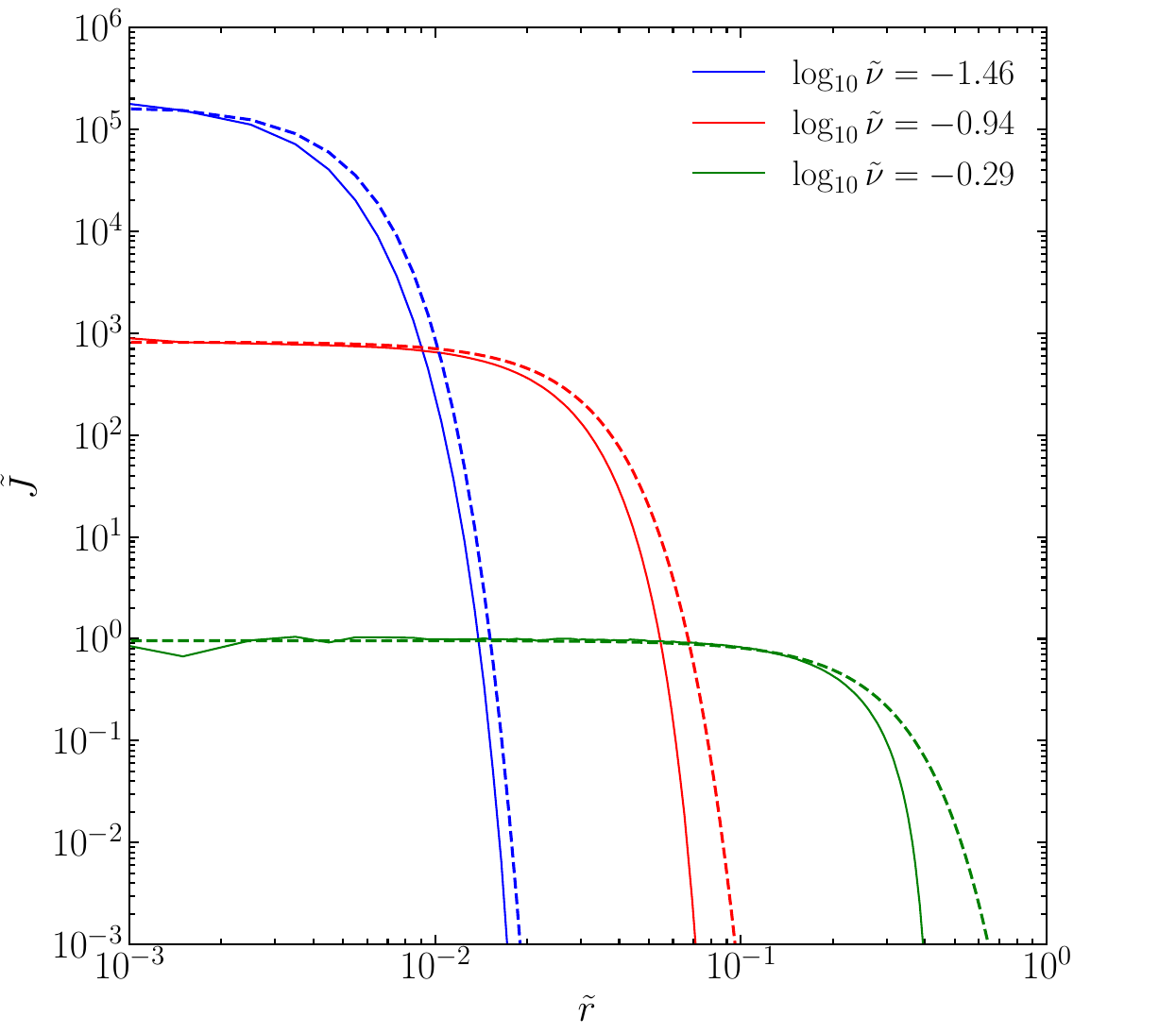}
\end{subfigure}
\caption{We demonstrate that our radiative transfer simulation reproduces the analytical solution by \citet{Loeb_1999} (equation~\ref{analytic}) for the specific intensity, $\tilde{J}=\tilde{J}(\tilde{r},\tilde{\nu})$, in a spherical medium of uniform number density of scatterers undergoing Hubble expansion at a temperature of $\SI{0}{\kelvin}$ (so that the line profile is given by a Lorentzian and the scatterers have no thermal motion). A single source emits at Ly$\alpha$ frequency from the centre of the sphere. Left panel: $\tilde{J}$ as a function of non-dimensionalised frequency, $\tilde{\nu}$, at fixed radii. Right panel: $\tilde{J}$ as a function of non-dimensionalised radius, $\tilde{r}$, at fixed frequencies. In both panels the solid curves show our results and the dashed curves show the analytical solution. See text for more details.}\label{fig:lr99test}
\end{figure*}

Figure~\ref{fig:lr99test} shows results from our general purpose MCRT code compared with the analytical solution, equation~\eqref{analytic}. We find reasonable agreement between the two.

\subsection{Scattering rate in a Hubble-expanding medium}\label{semelin}
For our final test we reproduce the $r^{-2.3}$ and $r^{-2}$ behaviour of the radial profile of $P_{\alpha}$ at small radii and large radii, respectively by a source which has a flat SED in a spherical medium of uniform hydrogen number density. See \citetalias{Zheng} for an explanation for this trend. We emphasise that $r^{-2.3}$ behaviour is seen only with the current SED type. Variations in the SED results in only slight deviations from $r^{-2.3}$ behaviour.

The following are our set-up details. We set the radius of the sphere to $100\,\mathrm{cMpc}$ and hydrogen number density at $\SI{254}{\metre^{-3}}$, which is the average value at $z=10$ for our cosmological parameters. We place the source at the centre with flat SED -- in terms of number of photons per unit frequency bin. Just as in the previous test we work at $\SI{0}{\kelvin}$ so that the line profile is a Lorentzian and gas atoms have no thermal motion. We used $10^7$ MC photons and 50 histogram bins for the radial profile.

\begin{figure}
\centering
\includegraphics[width=1\linewidth]{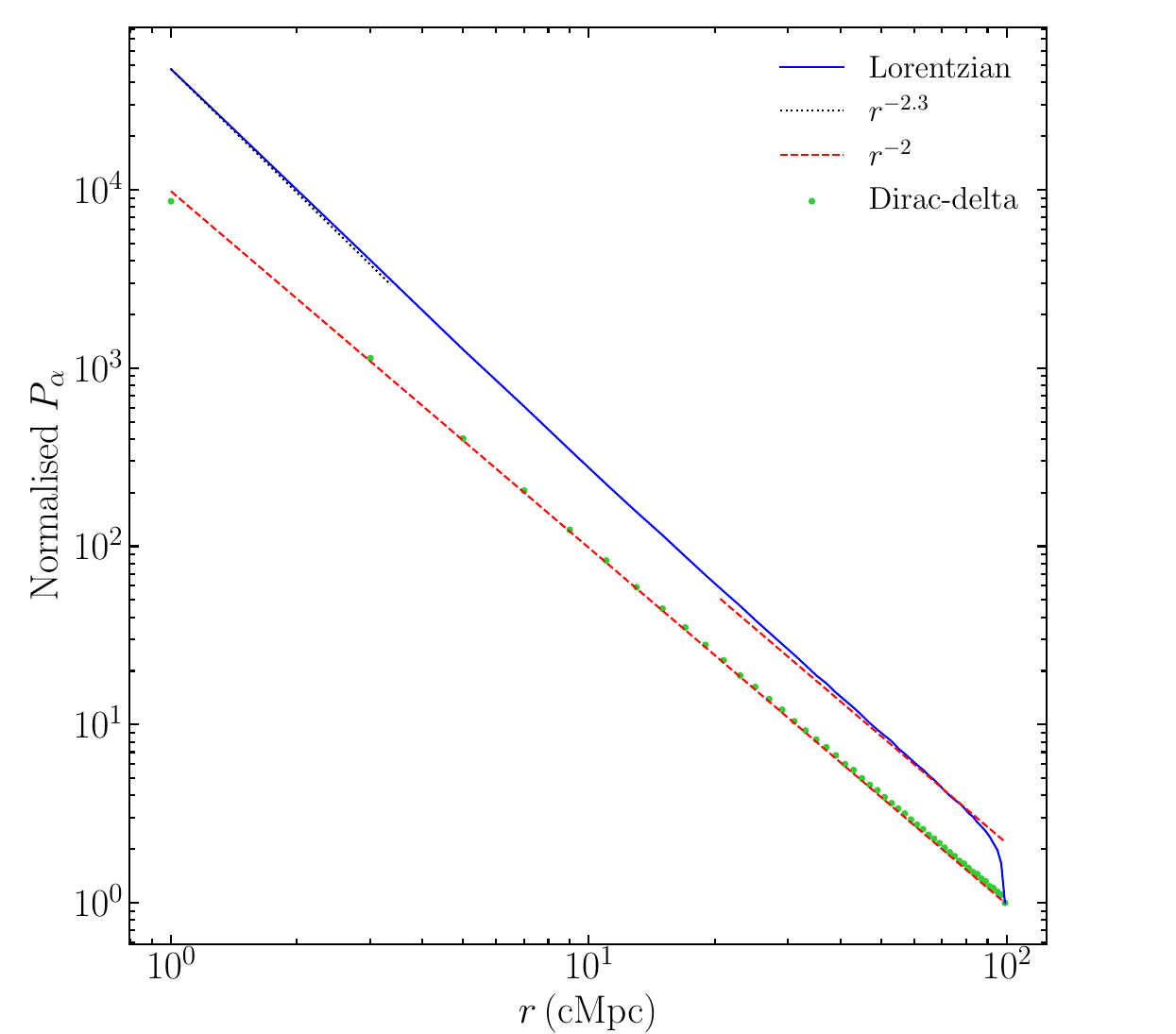}
\caption{Spherically-averaged profile of scattering rate $P_{\alpha}$. At small and large radius we recover the $r^{-2.3}$ and $r^{-2}$ behaviour, respectively when a Lorentzian line profile is used (blue-solid curve), consistent with previous literature. The medium is maintained at a uniform density and $\SI{0}{\kelvin}$ with a single source emitting from the centre with a flat SED in terms of number by frequency extending from Ly$\alpha$ to Ly$\beta$. We also recover the $r^{-2}$ behaviour (green circles) when the line profile is replaced by a Dirac-delta function. Red-dashed and black-dotted lines are guides representing slopes of $-2$ and $-2.3$ (on log-log scale), respectively. The different curves are separated only for clarity; the vertical positioning is unimportant.}\label{fig:semelin}
\end{figure}

We recover the $r^{-2.3}$ and $r^{-2}$ behaviour of $P_{\alpha}=P_{\alpha}(r)$ at small and large radii, respectively, in agreement with \citetalias{Semelin} and/or \citetalias{Zheng} as evident in Fig.~\ref{fig:semelin}. The blue-solid line shows the result from our simulation. (The minor deviation from $r^{-2}$ visible can be fixed by considering even larger radii). Just as for our third test run we show only a normalised version of $P_{\alpha}$ such that $P_{\alpha}=1$ at the farthest distance from the source.

As a further sanity check of our MCRT implementation we find that on setting a Dirac-delta\footnote{Actually we use the approximation $\ue^{-100x^2}$ for a delta-like function.} line profile we recover $r^{-2}$ trend of $P_{\alpha}(r)$ (green circles). As to why this should be the case can be seen as follows. A Dirac-delta line profile implies that multiple scatterings of Ly$\alpha$ are ignored. In this case the photon only scatters when it is exactly at the line centre frequency. Given an initial frequency one can compute the distance covered $r$ before they redshift and reach the line centre from the source. All photons of the same initial frequency will cross the sphere of radius $r$ at the same time and thus, the `Ly$\alpha$' flux and hence the scattering rate is simply proportional to $L/4\pi r^2$, where $L$ is the source luminosity \citep[see also][]{dijkstra_2008}. This implicit assumption for the calculation of background specific intensity has been used in 21-cm literature previously.

\section{Path-based vs point-based scattering rate}\label{pointvspath}
In this section we investigate two different styles of computing the scattering rate per atom, $P_\alpha$. In the first method we compute $P_\alpha$ by counting the number of scatterings in each cell while in the second method we sum the total optical depth traversed by all the photons in each cell. Throughout this work we have used the second definition even though the first one better aligns with the meaning of $P_\alpha$. The second version is an approximation to the first one and in general is accurate in regions of low optical depths \citep{Seon_2020}.

We follow the second definition because it gives a smoother version of $P_\alpha$, and hence $x_\alpha$. A photon will pass through several cells but scatter in a much smaller number of cells. First version will give non-zero $P_\alpha$ only in cells where scattering occurred while second version gives a non-zero $P_\alpha$ in all the cells through which the photon passed. Thus, the second version gives an `interpolated' map of the first version. As an example, for the toy model shown in Fig.~\ref{p_scheme} the first version of $P_\alpha$ would give 
$P_\alpha(c_1)=0$ and $$P_\alpha(c_2)=\frac{L/2}{V_{c_2}n_{\ion{H}{i},c_2}}\times2\,.$$This clearly shows that first version may lead to abrupt variations in $P_\alpha$ and the derived quantities.

\begin{figure}
\centering
\includegraphics[width=1\linewidth]{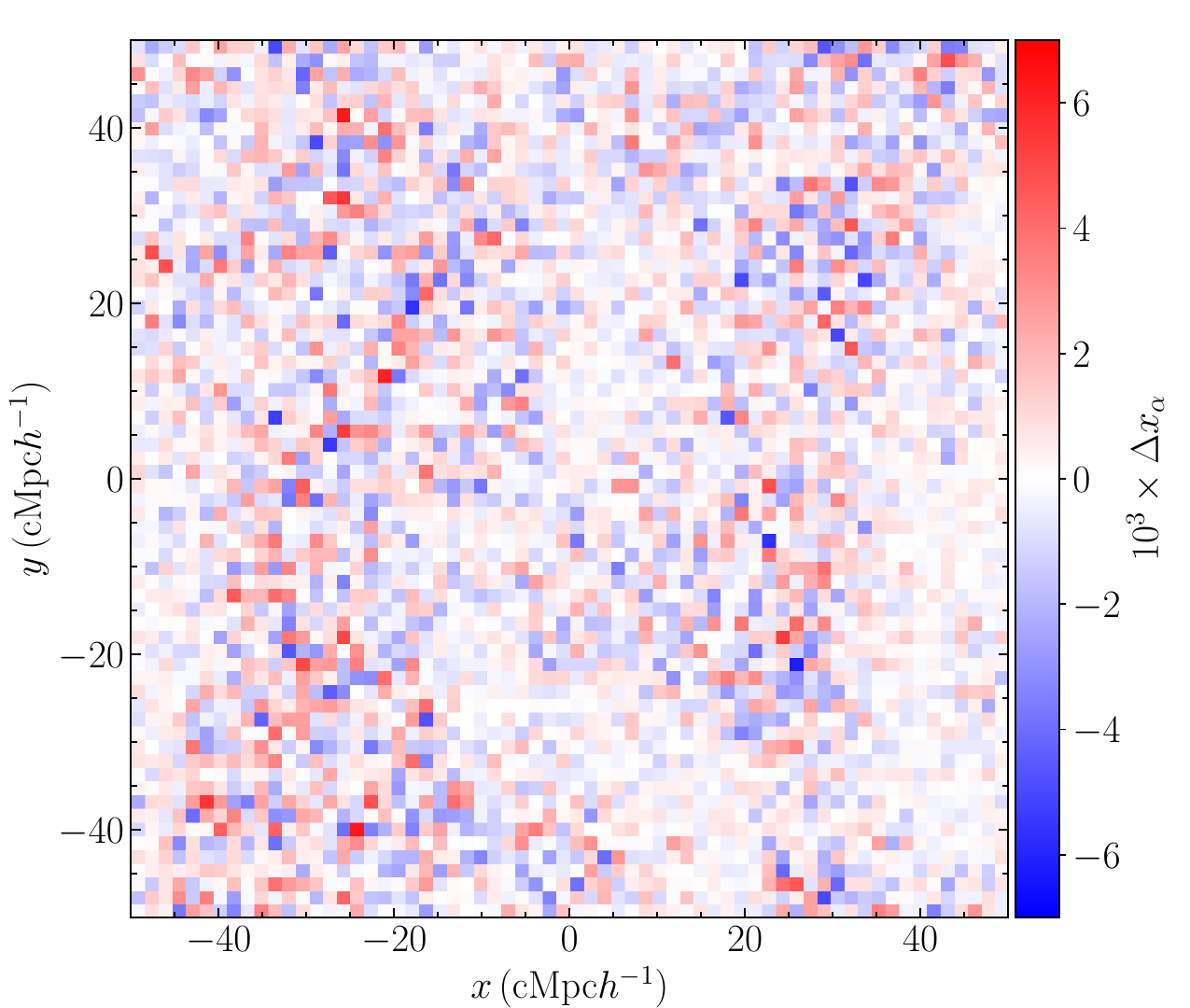}
\caption{Slice of difference between Ly$\alpha$ coupling through the middle of the box. The quantity $\Delta x_{\alpha} = (x_{\alpha,2}-x_{\alpha,1})$ represents the difference because of two different versions of $P_\alpha$, where $x_{\alpha,1,2}$ are the couplings in the $1^{\text{st}}$ and $2^{\text{nd}}$ version, respectively. In the $1^{\text{st}}$ version we count the number of scatterings in each cell while in the $2^{\text{nd}}$ version we sum the optical depths. We used $10^8$ MC photons for this result. As evident, overall the error is small and one may continue to work with the $2^{\text{nd}}$ version of $P_\alpha$.}\label{28vs29}
\end{figure}

When the number of MC photons is large the difference between the two versions is expected to be small as evident through Fig.~\ref{28vs29}. The mean and RMS of the difference over the complete domain are $\langle\Delta x_\alpha\rangle=3.6\times10^{-7}$ and $\sqrt{\langle(\Delta x_\alpha)^2\rangle}=1.5\times10^{-3}$, respectively. Since the overall error is small and leads to even smaller effect on the 21-cm signal we work with the second version of $P_\alpha$.


\bsp	
\label{lastpage}
\end{document}